\documentclass[aip,jmp,amsmath,amssymb,reprint]{revtex4}

\usepackage{graphicx}

\begin{document}

\newcommand\etal{\mbox{\textit{et al.}}}
\newcommand\Real{\mbox{Re}} 
\newcommand\Imag{\mbox{Im}} 
\newcommand\Rey{\mbox{\textit{Re}}} 
\newcommand\Pra{\mbox{\textit{Pr}}} 
\newcommand\Pec{\mbox{\textit{Pe}}} 

\title{Morphological instability of the solid-liquid interface in crystal growth under supercooled liquid film flow and natural convection airflow}

\author{Kazuto Ueno}

\email{k.ueno@kyudai.jp}


\author{Masoud Farzaneh}

\affiliation
{NSERC/Hydro-Quebec/UQAC Industrial Chair on Atmospheric Icing of Power Network Equipment (CIGELE)
and Canada Research Chair on Engineering of Power Network Atmospheric Icing (INGIVRE), 
Universit$\acute{e}$ du Qu$\acute{e}$bec $\grave{a}$ Chicoutimi, 
555 Boulevard de l'Universit$\acute{e}$, Chicoutimi, Qu$\acute{e}$bec G7H 2B1, Canada}


\begin{abstract}
Ring-like ripples on the surface of icicles are an example of morphological instability of the ice-water interface during ice growth under supercooled water film flow. The surface of icicles is typically covered with ripples of about 1 cm in wavelength, and the wavelength appears to be almost independent of external temperature, icicle radius, and volumetric water flow rate. One side of the water layer consists of the water-air surface and growing ice is the other. This is one of the more complicated moving phase boundary problems with two interfaces. 
A recent theoretical work [K. Ueno, Phys. Rev. E {\bf68}, 021603 (2003)] to address the underlying instability that produces ripples is based on the assumption of the absence of airflow around icicles. In this paper, we extend the previous theoretical framework to include a natural convection airflow ahead of the water-air surface and consider whether the effect of natural convection airflow on the wavelength of ripples produced on an ice surface is essential or not.
\end{abstract}


\maketitle

\section{Introduction}

Little is known on the study of morphological instability of the solid-liquid interface when a thin layer of moving fluid separates the developing solid from its surrounding. For example, the problem of icicle growth involves complex moving boundary problems with phase change.
When an icicle grows, a thin water film from the melting snow and ice at the root of the icicle flows down along its surface and refreezes onto it by releasing latent heat of solidification to the ambient air below 0 $^{\circ}$C. During the icicle growth, ice does not grow uniformly, but ring-like ripples are often observed on its surface. \cite{Maeno94} By supplying water continuously from the top of a wooden round stick and of a gutter on an inclined plane set in a cold room below 0 $^{\circ}$C, a ripple pattern similar to that observed on natural icicles is produced on the ice surface. \cite{Matsuda97}
Surprisingly, the distance between two peaks of ripples experimentally produced as well as that of natural icicles always measures around a centimeter scale. 

Theoretical works aimed at explaining the underlying dynamic instability that produces ripples are recent. \cite{Ogawa02, Ueno03, Ueno04, Ueno07, Ueno09} 
A stability analysis for the ice-water interface disturbance was developed based on heat flow in the water and atmosphere, and thin film water flow dynamics.
From the initial model, it was found that the ripple wavelength is determined from $\lambda=2\pi h_{0}\Pec_{l}/\alpha_{\rm max}$, and that the ripples should move down the icicle. \cite{Ogawa02} Here $h_{0}$ is the mean thickness of the water layer, $\Pec_{l}$ is the P${\rm \acute{e}}$clet number, which is a dimensionless number defined as the ratio of the heat transfer due to the water flow to that due to the thermal diffusion in the water layer, and $\alpha_{\rm max}$ is a dimensionless wave number at which the amplification rate of the ice-water interface disturbance acquires a maximum value. 
By considering different boundary conditions from those used in the initial model, a quite different ripple formation mechanism was developed. \cite{Ueno03, Ueno04} A new formula to determine the wavelength of ripples was derived: $\lambda=2\pi (a^{2}h_{0}\Pec_{l}/3)^{1/3}$, which contains two characteristic lengths $h_{0}$ and $a$. \cite{Ueno07} Here $a$ is the capillary length associated with the surface tension of the water-air surface. In the new model, the influence of the shape of the water-air surface on the growth condition of the ice-water interface was taken into account. Therefore, another length scale $a$ was introduced. The new model also predicted that ripples should move upward. The upward ripple translation was already suggested by the observation that many tiny air bubbles were trapped in the upper side of any protruded part of ripples during the icicle growth, and lined up upward. \cite{Maeno94} However, there was no theoretical explanation for the upward ripple translation mechanism.

Both models yield one-centimeter scale wavelength, but the translational direction of the ice ripples is opposite. Recently we solved numerically the same governing equations with the same boundary conditions as those used in the initial model. However, the numerically obtained amplification rate of the ice-water interface disturbances showed positive values for all wave numbers, \cite{Ueno09} which means that $\alpha_{\rm max}$ does not exist and there is no mechanism to select a characteristic length.
On the other hand, the analytical results for the amplification rate and the translation velocity of ice ripples obtained in the new model were in good agreement with those numerically calculated. Moreover, there was also good agreement between the theoretical predictions of the dependence of ripple wavelength on slope angles of the inclined plane and water supply rates and our experimental results. Finally, upward ripple motion at about half-speed of the mean growth rate of icicle radius was observed experimentally as theoretically predicted, but downward traveling ripples were not observed. \cite{Ueno09}

\begin{figure}
\begin{center}
\includegraphics[width=12cm,height=12cm,keepaspectratio,clip]{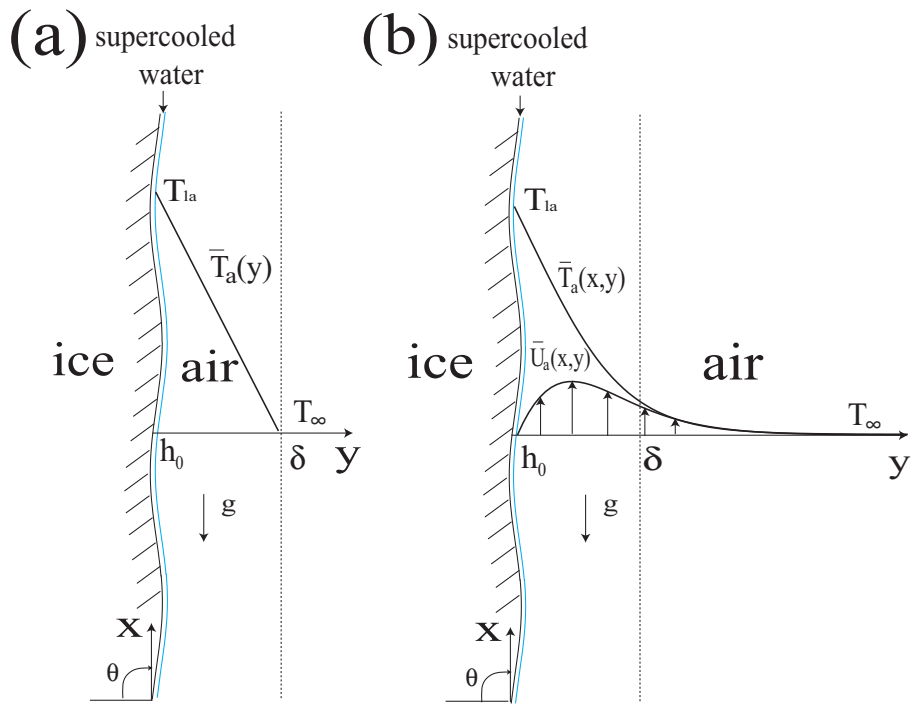}
\end{center}
\caption{Schematic view of the layer ahead of the water-air surface. Ice is covered with a supercooled water film. The $x$ axis is parallel to the direction of the supercooled water flow and the $y$ axis is normal to it. $T_{la}$ is the temperature at the water-air surface. (a) is the situation in absence of airflow. A linear air temperature distribution $\bar{T}_{a}(y)$ was assumed. (b) is the situation in presence of airflow. $\bar{U}_{a}(x,y)$ and $\bar{T}_{a}(x,y)$ are undisturbed velocity and temperature distributions. $h_{0}$ and $\delta$ are the thickness of the water layer and that of the thermal boundary layer, respectively. $g$ is the gravitational acceleration and $\theta$ is the angle with respect to the horizontal. The flowing supercooled water layer, not to scale, is much thinner than the thickness of the thermal boundary layer.}
\label{fig:ice-water-air}
\end{figure}
 
In the previous theoretical models, \cite{Ogawa02, Ueno03, Ueno04, Ueno07, Ueno09} ice was covered with a supercooled water layer and there was no airflow around icicles. The latent heat released at the ice-water interface was assumed to be transferred in the air by thermal diffusion through the water layer. As shown in Fig. \ref{fig:ice-water-air} (a), for simplicity, a linear air temperature distribution $\bar{T}_{a}(y)$ was assumed. \cite{Ueno04} From the energy conservation at the ice-water interface and water-air surface, the mean growth rate of icicle radius is given by $\bar{V}=-K_{a}T_{\infty}/(L\delta)$, where $K_{a}$ is the thermal conductivity of air, $L$ is the latent heat per unit volume and $T_{\infty}$ is the air temperature at a distance $\delta$ from the water-air surface. \cite{Ueno04}
The water layer of thickness $h_{0}$ changes by varying the water supply rate. \cite{Benjamin57, Landau59} However, since $\bar{V}$ does not include $h_{0}$, the icicle growth rate does not depend on the water supply rate. \cite{Maeno94, Ueno04} 
Since $\bar{V}$ contains the parameters $T_{\infty}$ and $\delta$, the ice growth rate is controlled by the rate of latent heat loss from the water-air surface to the surrounding air.   
However, it was not possible to estimate the value of $\bar{V}$ because the physical meaning of the assumed distance $\delta$ was unclear.

Recently, the growth of icicles has been treated as a free boundary problem to find an ideal growing shape for icicles. \cite{Short06} The latent heat transferred from the icicle surface to the surrounding air through the water layer leads to an increase in air temperature and to a change in density because it is temperature dependent. If the density decreases with increasing temperature, buoyancy force arises, and warmer air moves up along the ice surface. This effect is restricted to a thin layer ahead of the water-air surface, as shown in Fig. \ref{fig:ice-water-air} (b). 
Short $\etal$ emphasized the importance of heat transfer through such a convective boundary layer around icicles, and derived a formula for the ice growth velocity normal to the icicle's surface. The form is the same as $\bar{V}$ mentioned above, but a critical difference is that the length $\delta$ in paper \cite{Short06} is the boundary layer thickness. Hence, it was possible to estimate the values of $\delta$ and $\bar{V}$ if the value of an unknown parameter in $\delta$ was given. \cite{Short06, Ueno07} It was also suggested that similarity solutions for the coupled Navier-Stokes and heat transfer equations in the Boussinesq approximation can provide the basis for understanding of the boundary layer. \cite{Short06} In this paper, the value of the unknown parameter in $\delta$ is determined by obtaining the similarity solutions. 

Since heat transfer can be greatly influenced by the upward natural convection airflow, the question is whether the enhancement of heat transfer due to convection affects the wavelength of ripples on icicles.  
On the other hand, it is known that the wavelengths are almost independent of the length of the icicles and the ambient air temperature. 
In order to clarify these problems, in this paper, a linear stability analysis was performed on the ice-water interface disturbance during the ice growth in the presence of a supercooled water film flow and a natural convection airflow.

\section{Theory}

Instead of dealing with the elongated carrot-shaped geometry of the icicle, \cite{Short06} ice growth on a flat gutter on an inclined plane of finite length will be considered. The following theoretical analysis is restricted to two-dimensional vertical cross-sections of the gutter, as shown in Fig. \ref{fig:ice-water-air}. The origin of the $x$ axis is the bottom of the gutter and the $y$ axis is normal to it. 
What is new here is that the effect of a natural convection airflow is being incorporated into the previous theoretical frameworks \cite{Ueno03, Ueno04, Ueno07, Ueno09}  with modifications of some of boundary conditions, letting us treat synthetically heat flow in the ice, water and air through a disturbed ice-water interface and water-air surface, as well as thin water film flow and airflow.

\subsection{Governing equations}

The velocity components in the $x$ and $y$ directions in the water layer, $u_{l}$ and $v_{l}$, are governed by the Navier-Stokes equations driven by gravity and the continuity equation: \cite{Landau59}
\begin{equation}
\frac{\partial u_{l}}{\partial t}
+u_{l}\frac{\partial u_{l}}{\partial x}
+v_{l}\frac{\partial u_{l}}{\partial y} 
=-\frac{1}{\rho_{l}}\frac{\partial p_{l}}{\partial x}
+\nu_{l}\left(\frac{\partial^{2}u_{l}}{\partial x^{2}}
+\frac{\partial^{2}u_{l}}{\partial y^{2}}\right)-g\sin\theta,
\label{eq:geq-ul} 
\end{equation}

\begin{equation}
\frac{\partial v_{l}}{\partial t}
+u_{l}\frac{\partial v_{l}}{\partial x}
+v_{l}\frac{\partial v_{l}}{\partial y}
=-\frac{1}{\rho_{l}}\frac{\partial p_{l}}{\partial y}
+\nu_{l}\left(\frac{\partial^{2}v_{l}}{\partial x^{2}}
+\frac{\partial^{2}v_{l}}{\partial y^{2}}\right)-g\cos\theta, 
\label{eq:geq-vl}
\end{equation}

\begin{equation}
\frac{\partial u_{l}}{\partial x}+\frac{\partial v_{l}}{\partial y}=0,
\label{eq:continuity-water}
\end{equation}
where $\nu_{l}=1.8 \times 10^{-6}$ ${\rm m^{2}/s}$ and $\rho_{l}=1.0 \times 10^{3}$ ${\rm kg/m^{3}}$ are the kinematic viscosity and the density of water, $g$ the gravitational acceleration, $p_{l}$ the pressure in water. $\theta$ is the angle with respect to the horizontal, as shown in Fig. \ref{fig:ice-water-air}.

On the other hand, employing the Boussinesq approximation, the velocity components in the $x$ and $y$ directions in the air, $u_{a}$ and $v_{a}$, are governed by the following equations driven by buoyancy force and the continuity equation: \cite{Landau59} 
\begin{equation}
\frac{\partial u_{a}}{\partial t}
+u_{a}\frac{\partial u_{a}}{\partial x}
+v_{a}\frac{\partial u_{a}}{\partial y} 
=-\frac{1}{\rho_{\infty}}\frac{\partial (p_{a}-p_{a0})}{\partial x}
+\nu_{a}\left(\frac{\partial^{2}u_{a}}{\partial x^{2}}
+\frac{\partial^{2}u_{a}}{\partial y^{2}}\right)+g\beta(T_{a}-T_{\infty})\sin\theta,
\label{eq:geq-ua} 
\end{equation}

\begin{equation}
\frac{\partial v_{a}}{\partial t}
+u_{a}\frac{\partial v_{a}}{\partial x}
+v_{a}\frac{\partial v_{a}}{\partial y}
=-\frac{1}{\rho_{\infty}}\frac{\partial (p_{a}-p_{a0})}{\partial y}
+\nu_{a}\left(\frac{\partial^{2}v_{a}}{\partial x^{2}}
+\frac{\partial^{2}v_{a}}{\partial y^{2}}\right)+g\beta(T_{a}-T_{\infty})\cos\theta, 
\label{eq:geq-va}
\end{equation}

\begin{equation}
\frac{\partial u_{a}}{\partial x}+\frac{\partial v_{a}}{\partial y}=0,
\label{eq:continuity-air}
\end{equation}
where $p_{a}$ is the pressure in air, $p_{a0}$ the static pressure, $\rho_{\infty}$ the density of air at the temperature $T_{\infty}$, $\nu_{a}=1.3 \times 10^{-5}$ ${\rm m^{2}/s}$ and $\beta=3.7 \times 10^{-3}$ $K^{-1}$ are, respectively, the kinematic viscosity and the volumetric coefficient of thermal expansion for air. 
The continuity equations (\ref{eq:continuity-water}) and (\ref{eq:continuity-air}) can be satisfied by introducing the stream functions $\psi_{l}$ and $\psi_{a}$ such that
$u_{l}=\partial \psi_{l}/\partial y$, 
$v_{l}=-\partial \psi_{l}/\partial x$,
$u_{a}=\partial \psi_{a}/\partial y$ and  
$v_{a}=-\partial \psi_{a}/\partial x$.

Neglecting viscous dissipation in the energy equation, the equations for the temperatures in the ice $T_{s}$, water $T_{l}$ and air $T_{a}$ are \cite{Landau59}
\begin{equation}
\frac{\partial T_{s}}{\partial t}
=\kappa_{s}\left(\frac{\partial^{2} T_{s}}{\partial x^{2}}
+\frac{\partial^{2} T_{s}}{\partial y^{2}}\right),
\label{eq:geq-Ts}
\end{equation}

\begin{equation}
\frac{\partial T_{l}}{\partial t}
+u_{l}\frac{\partial T_{l}}{\partial x}
+v_{l}\frac{\partial T_{l}}{\partial y}
=\kappa_{l}\left(\frac{\partial^{2} T_{l}}{\partial x^{2}}+\frac{\partial^{2} T_{l}}{\partial y^{2}}\right),
\label{eq:geq-Tl}
\end{equation}

\begin{equation}
\frac{\partial T_{a}}{\partial t}
+u_{a}\frac{\partial T_{a}}{\partial x}
+v_{a}\frac{\partial T_{a}}{\partial y}
=\kappa_{a}\left(\frac{\partial^{2} T_{a}}{\partial x^{2}}
+\frac{\partial^{2} T_{a}}{\partial y^{2}}\right),
\label{eq:geq-Ta}
\end{equation}
where $\kappa_{s}=1.15 \times 10^{-6}$ ${\rm m^{2}/s}$, $\kappa_{l}=1.33 \times 10^{-7}$ ${\rm m^{2}/s}$ and $\kappa_{a}=1.87 \times 10^{-5}$ ${\rm m^{2}/s}$ are the thermal diffusivities of ice, water and air, respectively. Equations (\ref{eq:geq-ua}), (\ref{eq:geq-va}), (\ref{eq:continuity-air}) and (\ref{eq:geq-Ta}) are new part that has been added to the previous formulation. \cite{Ueno03, Ueno04, Ueno07, Ueno09}

\subsection{Boundary conditions at the ice-water interface and water-air surface}

\subsubsection{Hydrodynamic boundary conditions}

Neglecting the density difference between ice and water,
both velocity components $u_{l}$ and $v_{l}$ at a disturbed ice-water interface, $y=\zeta(t,x)$, must satisfy the no-slip condition:\cite{Ogawa02}
\begin{equation}
u_{l}|_{y=\zeta}=0,
\hspace{1cm}
v_{l}|_{y=\zeta}=0.
\label{eq:bc-ul-vl-zeta}
\end{equation}
The kinematic condition at a disturbed water-air surface, $y=\xi(t,x)$, is \cite{Benjamin57}
\begin{equation}
\frac{\partial \xi}{\partial t}+u_{l}|_{y=\xi}\frac{\partial \xi}{\partial x}=v_{l}|_{y=\xi}.
\label{eq:bc-kinematic-xi}
\end{equation}
The continuity of velocities of water film flow and airflow at the water-air surface is \cite{Yih67} 
\begin{equation}
u_{l}|_{y=\xi}=u_{a}|_{y=\xi},
\qquad
v_{l}|_{y=\xi}=v_{a}|_{y=\xi}.
\label{eq:bc-ul-vl-ua-va-xi}
\end{equation}
The condition for continuity of shear stress at the water-air surface is \cite{Craik66, Yih67}
\begin{equation}
\rho_{l}\nu_{l}\left(\frac{\partial u_{l}}{\partial y}\Big|_{y=\xi}
+\frac{\partial v_{l}}{\partial x}\Big|_{y=\xi}\right)
=\rho_{a}\nu_{a}\left(\frac{\partial u_{a}}{\partial y}\Big|_{y=\xi}
+\frac{\partial v_{a}}{\partial x}\Big|_{y=\xi}\right).
\label{eq:bc-shear-stress-xi}
\end{equation}
The difference of the normal stress on either side of the water-air surface must be the capillary force resisting displacement: \cite{Landau59, Craik66, Yih67}
\begin{equation}
-p_{a}|_{y=\xi}+2\rho_{a}\nu_{a}\frac{\partial v_{a}}{\partial y}\Big|_{y=\xi}
-\left(-p_{l}|_{y=\xi}+2\rho_{l}\nu_{l}\frac{\partial v_{l}}{\partial y}\Big|_{y=\xi}\right)
=-\gamma\frac{\partial^{2}\xi}{\partial x^2},
\label{eq:bc-normal-stress-xi}
\end{equation}
where $\gamma=7.6 \times 10^{-2}$ N/m is the surface tension of the water-air surface. The boundary conditions (\ref{eq:bc-ul-vl-zeta}) and (\ref{eq:bc-kinematic-xi}) are the same as those used in the previous papers. \cite{Ogawa02, Ueno03, Ueno04, Ueno07, Ueno09} Since an airflow is taken into account in this paper, the continuity condition of the water film and airflow velocities at the water-air surface is a new part, and the shear and normal stress conditions are modified from those in the previous papers. \cite{Ogawa02, Ueno03, Ueno04, Ueno07, Ueno09}   

\subsubsection{Thermodynamic boundary conditions}

The following boundary conditions are exactly the same as those in the previous papers. \cite{Ueno03, Ueno04, Ueno07, Ueno09}
The continuity condition of temperature is imposed at the ice-water interface:
\begin{equation}
T_{l}|_{y=\zeta}=T_{s}|_{y=\zeta}=T_{sl}+\Delta T_{sl},
\label{eq:Tsl}
\end{equation}
where $T_{sl}$ is the temperature at the flat ice-water interface and $\Delta T_{sl}$ is a deviation from it when the ice-water interface is disturbed.
The energy conservation at the ice-water interface is
\begin{equation}
L\left(\bar{V}+\frac{\partial \zeta}{\partial t} \right)
=K_{s}\frac{\partial T_{s}}{\partial y}\Big|_{y=\zeta}
      -K_{l}\frac{\partial T_{l}}{\partial y}\Big|_{y=\zeta},
\label{eq:heatflux-zeta}
\end{equation}
where $L=3.3 \times 10^{8}$ ${\rm J/m^{3}}$ is the latent heat per unit volume, and $K_{s}=2.22$ ${\rm J/(m\,K\,s)}$ and $K_{l}=0.56$ ${\rm J/(m\,K\,s)}$ are thermal conductivities of ice and water, respectively. 

The continuity condition of temperature is imposed at the water-air surface:
\begin{equation}
T_{l}|_{y=\xi}=T_{a}|_{y=\xi}=T_{la},
\label{eq:Tla}
\end{equation}
where $T_{la}$ is a temperature at the water-air surface.
The energy conservation at the water-air surface is
\begin{equation}
-K_{l}\frac{\partial T_{l}}{\partial y}\Big|_{y=\xi}
=-K_{a}\frac{\partial T_{a}}{\partial y}\Big|_{y=\xi},
\label{eq:heatflux-xi}
\end{equation}
where $K_{a}=0.024$ ${\rm J/(m\,K\,s)}$ is the thermal conductivity of air. 

In the initial model, \cite{Ogawa02} the continuity condition of temperature at the ice-water interface and water-air surface was 
$T_{s}|_{y=\zeta}=T_{l}|_{y=\zeta}=T_{sl}$ and $T_{l}|_{y=\xi}=T_{a}|_{y=\xi}=T_{la}+\Delta T_{la}$, where $\Delta T_{la}$ is a deviation from $T_{la}$ when the water-air surface is disturbed.
Instead, we use Eqs. (\ref{eq:Tsl}) and (\ref{eq:Tla}) as in the previous papers. \cite{Ueno03, Ueno04, Ueno07, Ueno09}
The difference in these boundary conditions led to critically different results between two models mentioned in the Introduction in this paper.  
When the chemical potential of water equals that of ice, it seems reasonable to assume the boundary condition $T_{s}|_{y=\zeta}=T_{l}|_{y=\zeta}=T_{sl}$ at the ice-water interface, then $T_{sl}$ is the equilibrium freezing temperature ($T_{sl}=0$ $^{\circ}$C for pure water). 
As in paper, \cite{Butler02} however, the chemical potential of water is not necessarily equal to that of ice because the ice-water coexistence considered here is expected to be in a non-equilibrium state in the presence of external disturbance at the water-air surface and shearing water flow. 
The deviation $\Delta T_{sl}$ at the ice-water interface caused by the external disturbance does not disappear by thermal diffusion in the water because the thermal relaxation time for the temperature fluctuation with about 1 cm corresponding to ice ripple wavelength is much longer than the time defined by the inverse of the shear rate of water film flow considered here. In other words, the equilibrium state at the ice-water interface is not attained in the presence of shearing water flow. \cite{Ueno09} We will see in \ref{sec:heatflux} that $\Delta T_{sl}$ is dependent on the temperature distribution in the water layer subject to the external disturbance at the water-air surface.
On the other hand, since shear stress has a value of zero at the water-air surface, the deviation $\Delta T_{la}$ at the water-air surface disappears by thermal diffusion in the air. Hence, the temperature at the water-air surface remains at $T_{la}$, which will be determined in \ref{sec:solutionTa}.

\subsection{Perturbation}

As shown in Fig. \ref{fig:ice-water-air}, only a one-dimensional perturbation in the $x$ direction of the ice-water interface with a small amplitude $\zeta_{k}$ is considered: $\zeta(t,x)=\zeta_{k}{\rm exp}[\sigma t+i kx]$,
where $k$ is the wave number and $\sigma=\sigma^{(r)}+i \sigma^{(i)}$. Here $\sigma^{(r)}$ and $v_{p} \equiv -\sigma^{(i)}/k$ are the amplification rate and the phase velocity of the perturbation, respectively. 
$\xi$, $\psi_{l}$, $\psi_{a}$, $p_{l}$, $p_{a}$, $T_{s}$, $T_{l}$ and $T_{a}$ are separated into unperturbed steady and perturbed parts as follows:
$\xi=h_{0}+\xi'$,
$\psi_{l}=\bar{\psi}_{l}+\psi'_{l}$,
$\psi_{a}=\bar{\psi}_{a}+\psi'_{a}$,
$p_{l}=\bar{P}_{l}+p'_{l}$,
$p_{a}=\bar{P}_{a}+p'_{a}$,
$T_{s}=\bar{T}_{s}+T'_{s}$,
$T_{l}=\bar{T}_{l}+T'_{l}$
and
$T_{a}=\bar{T}_{a}+T'_{a}$.
The corresponding perturbation of the water-air surface with a small amplitude $\xi_{k}$ is 
$\xi'(t,x)=\xi_{k}{\rm exp}[\sigma t+i kx]$.
As in the previous papers, \cite{Ueno03, Ueno04, Ueno07, Ueno09} the following calculation is based on a linear stability analysis taking into account only the first order of $\zeta_{k}$. The quasi-stationary approximation is also used: the time dependence of the perturbed part of equations can be neglected because the time evolution of the ice-water interface perturbation is considerably slow compared to that of the above perturbation fields. \cite{Caroli92}

\subsection{Equations of flow and temperature distributions in the air boundary layer}

Natural convection airflow considered here are restricted to a boundary layer regime and to conditions that lead to a similarity solution, that is, to a description of the flow by ordinary differential equations and boundary conditions in terms of a single coordinate $\eta(x,y)$. Under this assumption the unperturbed quantities $\bar{\psi}_{a}(x,y)$ and $\bar{T}_{a}(x,y)$ are expressed as follows:\cite{Gebhart73}
\begin{equation}
\bar{\psi}_{a}=u_{a0}\delta_{0}\bar{F}_{a}(\eta)=\nu_{a}Gr\bar{F}_{a}(\eta), \qquad
\bar{T}_{a*}=\frac{\bar{T}_{a}-T_{\infty}}{T_{la}-T_{\infty}},
\label{eq:basic-psi-T}
\end{equation}
where $\eta=(y-h_{0})/\delta_{0}$, $\delta_{0}=4x/Gr$ and $u_{a0}=\nu_{a}Gr^{2}/(4x)$. Here $Gr=4(Gr_{x}/4)^{1/4}$ is the modified local Grashof number, $Gr_{x}=g\beta\Delta T_{a}x^{3}/\nu_{a}^{2}$ being the local Grashof number. $\Delta T_{a}=T_{la}-T_{\infty}$ is the temperature difference between the water-air surface and the ambient air temperature far away. $x$ is the distance from the bottom of the gutter. 

Applying the boundary layer approximation to the Boussinesq equations (\ref{eq:geq-ua}), (\ref{eq:geq-va}), (\ref{eq:continuity-air}) and  (\ref{eq:geq-Ta}), $\bar{\psi}_{a}(x,y)$ and $\bar{T}_{a}(x,y)$ are governed by \cite{Landau59, Schlichting99}  
\begin{equation}
\frac{\partial \bar{\psi}_{a}}{\partial y}\frac{\partial^{2}\bar{\psi}_{a}}{\partial x \partial y}
-\frac{\partial \bar{\psi}_{a}}{\partial x}\frac{\partial^{2} \bar{\psi}_{a}}{\partial y^{2}} 
=\nu_{a}\frac{\partial^{3}\bar{\psi}_{a}}{\partial y^{3}}+g\beta(\bar{T}_{a}-T_{\infty})\sin\theta,
\label{eq:geq-ua-basic} 
\end{equation}
\begin{equation}
\frac{\partial \bar{\psi}_{a}}{\partial y}\frac{\partial \bar{T}_{a}}{\partial x}
-\frac{\partial \bar{\psi}_{a}}{\partial x}\frac{\partial \bar{T}_{a}}{\partial y}
=\kappa_{a}\frac{\partial^{2} \bar{T}_{a}}{\partial y^{2}}.
\label{eq:geq-Ta-basic}
\end{equation}
When Eq. (\ref{eq:basic-psi-T}) is substituted into Eqs. (\ref{eq:geq-ua-basic}) and  (\ref{eq:geq-Ta-basic}), the dimensionless functions $\bar{F}_{a}$ and $\bar{T}_{a*}$ are obtained from the two coupled ordinary differential equations: \cite{Landau59} 
\begin{equation}
\frac{d^{3}\bar{F}_{a}}{d\eta^{3}}
=-3\bar{F}_{a}\frac{d^{2}\bar{F}_{a}}{d\eta^{2}}
+2\left(\frac{d\bar{F}_{a}}{d\eta}\right)^{2}-\bar{T}_{a*}\sin\theta,
\label{eq:geq-basicFa}
\end{equation}
\begin{equation}
\frac{d^{2}\bar{T}_{a*}}{d\eta^{2}}
=-3Pr_{a}\bar{F}_{a}\frac{d\bar{T}_{a*}}{d\eta},
\label{eq:geq-basicTa}
\end{equation}
where $Pr_{a}=\nu_{a}/\kappa_{a}=0.7$ is the Prandtl number of air. 

We assume stream function disturbance $\psi'_{a}$ and temperature disturbance $T'_{a}$ in the air to be of the form: 
\begin{equation}
\psi'_{a}=u_{a0}f_{a}(\eta)\xi_{k}{\rm exp}[\sigma t +ikx], \qquad
T'_{a}=H_{a}(\eta)\bar{G}_{a}\xi_{k}{\rm exp}[\sigma t +ikx], 
\label{eq:pert-psi-T}
\end{equation}
where $f_{a}$ and $H_{a}$ are the dimensionless disturbance amplitude functions, and $\bar{G}_{a} \equiv -\partial \bar{T}_{a}/\partial y|_{y=h_{0}}$.
When $\psi_{a}=\bar{\psi}_{a}+\psi'_{a}$ and $T_{a}=\bar{T}_{a}+T'_{a}$ are substituted into the complete equations (\ref{eq:geq-ua}), (\ref{eq:geq-va}) and  (\ref{eq:geq-Ta}), we obtain the differential equations for the functions $f_{a}$ and $H_{a}$:

\begin{eqnarray}
\frac{d^{4}f_{a}}{d\eta^4}
&=&-3\bar{F}_{a}\frac{d^{3}f_{a}}{d\eta^{3}}
+\left(2\mu_{a}^{2}+i\mu_{a}Gr\frac{d\bar{F}_{a}}{d\eta}\right)\frac{d^{2}f_{a}}{d\eta^{2}}
+\left\{\mu_{a}^{2}\left(3\bar{F}_{a}+2\eta\frac{d\bar{F}_{a}}{d\eta}\right) 
+\frac{d^{2}\bar{F}_{a}}{d\eta^{2}}\right\}\frac{df_{a}}{d\eta}\nonumber \\
&& -\left\{\mu_{a}^{4}+\mu_{a}^{2}(6+i\mu_{a}Gr)\frac{d\bar{F}_{a}}{d\eta}
+(2+i\mu_{a}Gr)\frac{d^{3}\bar{F}_{a}}{d\eta^{3}}\right\}f_{a} \nonumber \\
&& -\bar{G}_{a*}\frac{dH_{a}}{d\eta}\sin\theta+i\mu_{a}\bar{G}_{a*}H_{a}\cos\theta,
\label{eq:geq-fa} 
\end{eqnarray}
\begin{eqnarray}
\frac{d^{2}H_{a}}{d\eta^{2}}
&=&-3Pr_{a}\bar{F}_{a}\frac{dH_{a}}{d\eta}
+\left\{\mu_{a}^{2}+Pr_{a}(-1+i\mu_{a}Gr)\frac{d\bar{F}_{a}}{d\eta}\right\}H_{a}\nonumber \\  
&&-Pr_{a}/\bar{G}_{a*}(2+i\mu_{a}Gr)\frac{d\bar{T}_{a*}}{d\eta}f_{a},
\label{eq:geq-Ha}
\end{eqnarray}
where $\mu_{a}=k\delta_{0}$ is the dimensionless wave number normalized by the length $\delta_{0}$, and $\bar{G}_{a*} \equiv -d\bar{T}_{a*}/d\eta|_{\eta=0}$, whose value depends on the Prandtl number. In the stability analysis, \cite{Gebhart73} $\bar{v}_{a}=-\partial\bar{\psi}_{a}/\partial x$ and $\partial\bar{T}_{a*}/\partial x$ were neglected because the derivatives of the unperturbed fields quantities  $\bar{F}_{a}$ and $\bar{T}_{a*}$ with respect to $x$ were assumed to be much smaller than those with respect to $y$. In this paper, however, these quantities in Eqs. (\ref{eq:geq-fa}) and (\ref{eq:geq-Ha}) are retained because if we neglect them, $\sigma^{(r)}$ and $v_{p}$ do not converge zero as $\mu_{a}$ approaches zero.

\subsection{Equations of flow and temperature distributions in the water layer}

The stream function disturbance $\psi'_{l}$ and temperature disturbance $T'_{l}$ in the water layer are assumed to be of the form: \cite{Ueno03, Ueno04, Ueno07, Ueno09}
\begin{equation}
\psi'_{l}=u_{l0}f_{l}(y_{*})\zeta_{k}{\rm exp}[\sigma t +ikx], \qquad
T'_{l}=H_{l}(y_{*})\bar{G}_{l}\zeta_{k}{\rm exp}[\sigma t +ikx], 
\label{eq:pert-psil-Tl}
\end{equation}
where $y_{*}=y/h_{0}$, and $f_{l}$ and $H_{l}$ are the dimensionless disturbance amplitude functions. It is also assumed that the unperturbed temperature distribution in the water layer is linear, then $\bar{G}_{l} \equiv -\partial \bar{T}_{l}/\partial y|_{y=h_{0}}=(T_{sl}-T_{la})/h_{0}$. 

When $\psi_{l}=\bar{\psi}_{l}+\psi'_{l}$ is substituted into Eqs. (\ref{eq:geq-ul}) and (\ref{eq:geq-vl}), the perturbed part yields the following Orr-Sommerfeld equation for $f_{l}$: \cite{Ueno03, Ueno09}
\begin{equation}
\frac{d^{4}f_{l}}{dy_{*}^{4}}
=\left(2\mu_{l}^{2}+i\mu_{l} \Rey_{l}\bar{U}_{l*}\right)\frac{d^{2}f_{l}}{dy_{*}^{2}}
-\left\{\mu_{l}^{4}+i\mu_{l} \Rey_{l}\left(\mu_{l}^{2}\bar{U}_{l*}+\frac{d^{2}\bar{U}_{l*}}{dy_{*}^{2}}\right)\right\}f_{l},
\label{eq:geq-fl}
\end{equation}
where $\mu_{l}=kh_{0}$ is the dimensionless wave number normalized by the length $h_{0}$, $\bar{U}_{l*}(y_{*})$ is the dimensionless velocity distribution in the water layer in the unperturbed state, and $\Rey_{l}\equiv u_{l0}h_{0}/\nu_{l}=3Q/(2l\nu_{l})$ is the Reynolds number. Here $Q/l$ is the water supply rate per width.

When $T_{l}=\bar{T}_{l}+T'_{l}$ are substituted into (\ref{eq:geq-Tl}), the perturbed part yields the equation for $H_{l}$: \cite{Ueno03, Ueno09}
\begin{equation}
\frac{d^{2}H_{l}}{dy_{*}^{2}}
=(\mu_{l}^{2}+i\mu_{l} \Pec_{l}\bar{U}_{l*})H_{l} 
-i\mu_{l} \Pec_{l}\frac{d\bar{T}_{l*}}{d y_{*}}f_{l},
\label{eq:geq-Hl}
\end{equation}
where $\bar{T}_{l*}(y_{*})\equiv (\bar{T}_{l}(y_{*})-T_{sl})/(T_{sl}-T_{la})=-y_{*}$ is the dimensionless temperature distribution in the water layer in the unperturbed state, and $\Pec_{l}\equiv u_{l0}h_{0}/\kappa_{l}=3Q/(2l\kappa_{l})$ is the ${\rm P\acute{e}clet}$ number.

\subsection{\label{sec:linearization}Linearization of boundary conditions}

First, linearizing Eq. (\ref{eq:bc-ul-vl-zeta}) at $y=0$ yields, to the first order in $\zeta_{k}$,  
\begin{equation}
\frac{df_{l}}{dy_{*}}\Big|_{y_{*}=0}+\frac{d\bar{U}_{l*}}{dy_{*}}\Big|_{y_{*}=0}=0,  \qquad
f_{l}|_{y_{*}=0}=0,
\label{eq:ul-vl-h0}
\end{equation}
From the linearization of Eq. (\ref{eq:bc-kinematic-xi}) at $y=h_{0}$, the relation between the amplitude of the water-air surface and that of the ice-water interface is obtained: $\xi_{k}=-(f_{l}|_{y_{*}=1}/\bar{U}_{l*}|_{y_{*}=1})\zeta_{k}$. \cite{Ueno03, Ueno04, Ueno07, Ueno09}  

Second, linearizing Eq. (\ref{eq:bc-ul-vl-ua-va-xi}) at $y=h_{0}$ yields, to the zeroth order in $\xi_{k}$, 
\begin{equation}
\frac{d\bar{F}_{a}}{d\eta}\Big|_{\eta=0}=\frac{u_{l0}}{u_{a0}}\bar{U}_{l*}|_{y_{*}=1}, 
\qquad 
\bar{F}_{a}|_{\eta=0}=0,
\label{eq:ua-va-h0}
\end{equation}
and to the first order in $\xi_{k}$, 
\begin{eqnarray}
\frac{df_{a}}{d\eta}\Big|_{\eta=0}=-\frac{d^{2}\bar{F}_{a}}{d\eta^{2}}\Big|_{\eta=0}
+\frac{\delta_{0}}{h_{0}}\frac{u_{l0}}{u_{a0}}
\left\{\frac{d\bar{U}_{l*}}{dy_{*}}\Big|_{y_{*}=1}-\left(\frac{df_{l}}{dy_{*}}\Big|_{y_{*}=1}\Big/f_{l}|_{y_{*}=1}\right)
\bar{U}_{l*}|_{y_{*}=1}\right\},
\nonumber \\
f_{a}|_{\eta=0}=-\frac{u_{l0}}{u_{a0}}\bar{U}_{l*}|_{y_{*}=1}.
\label{eq:ua-va-xi}
\end{eqnarray}
The values of $u_{a0}$ are 0.38 m/s at $x=0.1$ m and 1.2 m/s at $x=1.0$ m for $\Delta T_{a}=10$ $^{\circ}$C. 
On the other hand, the surface velocity of the water layer,
$u_{l0}=[g\sin\theta/(2\nu_{l})]^{1/3}[3Q/(2l)]^{2/3}$,
is about $0.78 \sim 3.62$ cm/s for typical values of $Q/l=10 \sim 100$ [(ml/h)/cm] and $\theta=\pi/2$.  It should be noted that the velocity of the water film flow is much less than that of airflow. Therefore, the first equation in (\ref{eq:ua-va-h0}) and the second equation in (\ref{eq:ua-va-xi}) are approximated as $d\bar{F}_{a}/d\eta|_{\eta=0}=0$ and $f_{a}|_{\eta=0}=0$, respectively. 
Even though a thin fluid layer of water flows down the ice surface, the no-slip condition at the water-air surface of the flowing water film is nearly satisfied for the velocities: 
$\bar{u}_{a}=\partial \bar{\psi}_{a}/\partial y$, 
$\bar{v}_{a}=-\partial \bar{\psi}_{a}/\partial x$ and
$v'_{a}=-\partial \psi'_{a}/\partial x$. 
The values of $\delta_{0}$ are 3.7 mm at $x=0.1$ m and 6.6 mm at $x=1.0$ m for $\Delta T_{a}=10$ $^{\circ}$C. 
On the other hand, the mean thickness of the water layer, 
$h_{0}=[3\nu_{l}/(g\sin\theta)Q/l]^{1/3}$,
is about $53 \sim 115$ $\mu$m for values of $Q/l=10 \sim 100$ [(ml/h)/cm] and $\theta=\pi/2$. 
Since $\delta_{0}/h_{0} \gg 1$, the second term on the right hand side of the first equation in Eq. (\ref{eq:ua-va-xi}) cannot be neglected. Hence, the no-slip condition cannot be applied to 
$u'_{a}=\partial \psi'_{a}/\partial y$
at the water-air surface of the flowing water film. 

Third, linearizing Eq. (\ref{eq:bc-shear-stress-xi}) at $y=h_{0}$ yields, to the zeroth order in $\xi_{k}$, 
\begin{equation}
\frac{d\bar{U}_{l*}}{dy_{*}}\Big|_{y_{*}=1}
=\frac{\rho_{a}\nu_{a}(u_{a0}d^{2}\bar{F}_{a}/d\eta^{2}|_{\eta=0})/\delta_{0}}{\rho_{l}\nu_{l}u_{l0}/h_{0}}
\equiv R_{\tau_{al}},
\label{eq:shear-stress-h0}
\end{equation}
and to the first order in $\xi_{k}$, 
\begin{eqnarray}
\frac{d^{2}f_{l}}{dy_{*}^{2}}\Big|_{y_{*}=1}
+\left(-\frac{d^{2}\bar{U}_{l*}}{dy_{*}^{2}}\Big|_{y_{*}=1}\Big/\bar{U}_{l*}|_{y_{*}=1}+\mu_{l}^{2}\right)f_{l}|_{y_{*}=1}
\nonumber \\
=-\frac{\rho_{a}\nu_{a}}{\rho_{l}\nu_{l}}\left(\frac{h_{0}}{\delta_{0}}\right)^{2}\frac{u_{a0}}{u_{l0}}
\left\{\frac{d^{2}f_{a}}{d\eta^{2}}\Big|_{\eta=0}+\frac{d^{3}\bar{F}_{a}}{d\eta^{3}}\Big|_{\eta=0}+\mu_{a}^{2}f_{a}|_{\eta=0}\right\}
f_{l}|_{y_{*}=1}/\bar{U}_{l*}|_{y_{*}=1},
\label{eq:shear-stress-xi}
\end{eqnarray}
where $R_{\tau_{al}}$ on the right hand side of Eq. (\ref{eq:shear-stress-h0}) should be nearly considered as the ratio of the shear stress of airflow at the water-air surface to that of the water film flow at the ice-water interface. 
It is assumed that
$p'_{l}=\rho_{l}u_{l0}^{2}\Pi_{l}(y_{*})\zeta_{k}/h_{0}{\rm exp}[\sigma t+i kx]$ and 
$p'_{a}=\rho_{a}u_{a0}^{2}\Pi_{a}(\eta)\xi_{k}/\delta_{0}{\rm exp}[\sigma t+i kx]$, where $\Pi_{l}$ and $\Pi_{a}$ are dimensionless amplitudes.
Substituting these forms into Eq. (\ref{eq:bc-normal-stress-xi}) and linearizing them at $y=h_{0}$ yields, to the first order in $\xi_{k}$, 
\begin{eqnarray}
\frac{d^{3}f_{l}}{dy_{*}^{3}}\Big|_{y_{*}=1}
-(i\mu_{l} \Rey_{l}\bar{U}_{l*}|_{y_{*}=1}+3\mu_{l}^{2})\frac{df_{l}}{dy_{*}}\Big|_{y_{*}=1}
+i\left(\mu_{l}\Rey_{l}\frac{d\bar{U}_{l*}}{dy_{*}}\Big|_{y_{*}=1}+\alpha/\bar{U}_{l*}|_{y_{*}=1}\right)f_{l}|_{y_{*}=1}
\nonumber \\
=-\frac{\rho_{a}\nu_{a}}{\rho_{l}\nu_{l}}\left(\frac{h_{0}}{\delta_{0}}\right)^{3}\frac{u_{a0}}{u_{l0}}
\left\{
\frac{d^{3}f_{a}}{d\eta^{3}}\Big|_{\eta=0}
-\left(i\mu_{a}Gr\frac{d\bar{F}_{a}}{d\eta}\Big|_{\eta=0}+3\mu_{a}^{2}\right)\frac{df_{a}}{d\eta}\Big|_{\eta=0} \right.
\nonumber \\
\left.
+i\mu_{a}Gr\frac{d^{2}\bar{F}_{a}}{d\eta^{2}}\Big|_{\eta=0}f_{a}|_{\eta=0}+\bar{G}_{a*}H_{a}|_{\eta=0}\sin\theta
\right\}
f_{l}|_{y_{*}=1}/\bar{U}_{l*}|_{y_{*}=1},
\label{eq:normal-stress-xi}
\end{eqnarray}
where 
\begin{equation}
\alpha=2(\cot\theta)\mu_{l}+\frac{2}{\sin\theta}\left(\frac{a}{h_{0}}\right)^{2}\mu_{l}^{3},
\label{eq:alpha}
\end{equation}
represents a parameter relevant to the restoring force due to the surface tension and gravity acting on the water-air surface. \cite{Ueno03, Benjamin57} Here $a=[\gamma/(\rho_{l}g)]^{1/2}$ is the capillary length associated with the surface tension $\gamma$ of the water-air surface. \cite{Landau59}

From the boundary condition (\ref{eq:shear-stress-h0}) and the no-slip condition $\bar{U}_{l*}|_{y_{*}=0}=0$, the velocity profile in the water layer is given by $\bar{U}_{l*}=y_{*}^{2}+(R_{\tau_{al}}-2)y_{*}$. However, $R_{\tau_{al}}$ is extremely small because the ratio of the viscosity of air to that of water,
$\rho_{a}\nu_{a}/\rho_{l}\nu_{l}$, 
as well as $h_{0}/\delta_{0}$ are much smaller than 1.  
Therefore, the shear stress-free condition, $d\bar{U}_{l*}/dy_{*}|_{y{*}=1}=0$, holds at the unperturbed water-air surface. Thus the velocity profile in the water layer is still the half-parabolic form, $\bar{U}_{l*}=y_{*}^{2}-2y_{*}$, so that the values of $\bar{U}_{l*}|_{y_{*}=1}=-1$, $d\bar{U}_{l*}/dy_{*}|_{y_{*}=0}=-2$ and $d^{2}\bar{U}_{l*}/dy_{*}^{2}|_{y_{*}=1}=2$ are used in the above boundary conditions. 
Similarly, since $\rho_{a}\nu_{a}/\rho_{l}\nu_{l} \ll 1$ and $h_{0}/\delta_{0} \ll 1$ on the right hand side of Eqs. (\ref{eq:shear-stress-xi}) and (\ref{eq:normal-stress-xi}), the influence of the perturbed part of shear and normal stresses due to airflow on the water film flow at the water-air surface is negligible. Therefore, the boundary conditions for the shear and normal stresses at the perturbed water-air surface become the same as those used in the previous papers. \cite{Ueno03, Ueno04, Ueno07, Ueno09} 

Finally, linearizing Eq. (\ref{eq:Tla}) at $y=h_{0}$ yields, to the zeroth order in $\xi_{k}$, 
$\bar{T}_{l*}|_{y_{*}=1}=-1$, 
$\bar{T}_{a*}|_{\eta=0}=1$,
and to the first order in $\xi_{k}$, 
\begin{equation}
H_{l}|_{y_{*}=1}+f_{l}|_{y_{*}=1}/\bar{U}_{l*}|_{y_{*}=1}=0, \qquad
H_{a}|_{\eta=0}=1.
\label{eq:Tla-xi}
\end{equation}
Linearizing Eq. (\ref{eq:heatflux-xi}) at $y=h_{0}$ yields, to the first order in $\xi_{k}$, 
\begin{equation}
\frac{dH_{l}}{dy_{*}}\Big|_{y_{*}=1}
-\frac{h_{0}}{\delta_{0}}\left(-\frac{dH_{a}}{d\eta}\Big|_{\eta=0}\right)f_{l}|_{y_{*}=1}/\bar{U}_{l*}|_{y_{*}=1}=0.
\label{eq:heatflux-xi-h0}
\end{equation}
It is convenient to define 
\begin{equation}
G'^{(r)}_{a}\equiv \frac{h_{0}}{\delta_{0}}\left(-\frac{dH_{a}^{(r)}}{d\eta}\Big|_{\eta=0}\right),\qquad
G'^{(i)}_{a}\equiv \frac{h_{0}}{\delta_{0}}\left(-\frac{dH_{a}^{(i)}}{d\eta}\Big|_{\eta=0}\right),
\label{eq:Gar-Gai}  
\end{equation}
which represents the real and imaginary parts of the perturbed part of the air temperature gradient at the water-air surface. 
It should be noted that Eq. (\ref{eq:geq-fl}) can be independently solved with the boundary conditions (\ref{eq:ul-vl-h0}), (\ref{eq:shear-stress-xi}) and (\ref{eq:normal-stress-xi}) without considering the influence of airflow. Therefore, $f_{l}$ in Eqs. (\ref{eq:Tla-xi}) and (\ref{eq:heatflux-xi-h0}) is the same form as that in the absence of airflow.
The perturbed part of temperature in the water layer is affected by the airflow through the perturbed part of the air temperature gradient in Eq. (\ref{eq:heatflux-xi-h0}). 

\subsection{Dispersion relation}

From the perturbed part of Eqs. (\ref{eq:Tsl}) and (\ref{eq:heatflux-zeta}), the dispersion relation for the perturbation of the ice-water interface is given by \cite{Ueno03, Ueno04, Ueno07, Ueno09}
\begin{equation}
\sigma=\frac{\bar{V}}{h_{0}}\left\{-\frac{dH_{l}}{dy_{*}}\Big|_{y_{*}=0}+K^{s}_{l}\mu_{l} (H_{l}|_{y_{*}=0}-1)\right\},
\label{eq:dispersion}
\end{equation}
where $K^{s}_{l}=K_{s}/K_{l}=3.96$ is the ratio of the thermal conductivity of ice to that of water. The real and imaginary parts of Eq. (\ref{eq:dispersion}) give the dimensionless amplification rate $\sigma_{*}^{(r)}\equiv \sigma^{(r)}/(\bar{V}/h_{0})$ and the dimensionless phase velocity $v_{p*}\equiv -\sigma^{(i)}/(k\bar{V})$, respectively, 
\begin{equation}
\sigma_{*}^{(r)}=-\frac{dH_{l}^{(r)}}{dy_{*}}\Big|_{y_{*}=0}+K^{s}_{l}\mu_{l}(H_{l}^{(r)}|_{y_{*}=0}-1),
\label{eq:amplificationrate}
\end{equation}
\begin{equation}
v_{p*}=-\frac{1}{\mu_{l}}\left(-\frac{dH_{l}^{(i)}}{dy_{*}}\Big|_{y_{*}=0}+K^{s}_{l}\mu_{l}H_{l}^{(i)}|_{y_{*}=0}\right),
\label{eq:phasevelocity}
\end{equation}
where $H_{l}^{(r)}$ and $H_{l}^{(i)}$ are the real and imaginary parts of $H_{l}$.

The numerical procedure for obtaining the wavelength and phase velocity of ice ripples is as follows.
First, Eq. (\ref{eq:geq-fl}) is solved with the boundary conditions (\ref{eq:ul-vl-h0}), (\ref{eq:shear-stress-xi}) and (\ref{eq:normal-stress-xi}). Substituting the obtained solution $f_{l}$ into Eq. (\ref{eq:ua-va-xi}), then Eqs. (\ref{eq:geq-basicFa}), (\ref{eq:geq-basicTa}), (\ref{eq:geq-fa}) and (\ref{eq:geq-Ha}) must be solved simultaneously for a given $Gr$ with the following boundary conditions: 
Eq. (\ref{eq:ua-va-h0}), 
$d\bar{F}_{a}/d\eta|_{\eta=\infty}=0$,
$\bar{T}_{a*}|_{\eta=0}=1$, 
$\bar{T}_{a*}|_{\eta=\infty}=0$,
Eq. (\ref{eq:ua-va-xi}), 
$df_{a}/d\eta|_{\eta=\infty}=0$, 
$f_{a}|_{\eta=\infty}=0$,  
$H_{a}|_{\eta=0}=1$ and 
$H_{a}|_{\eta=\infty}=0$.
Here, it is assumed that 
$u'_{a}|_{y=\infty}=\partial \psi'_{a}/\partial y|_{y=\infty}=0$, 
$v'_{a}|_{y=\infty}=-\partial \psi'_{a}/\partial x|_{y=\infty}=0$, and
$T_{a}'|_{y=\infty}=0$. \cite{Gebhart73} 
Substituting the obtained solutions $f_{l}$ and $H_{a}$ into the boundary conditions (\ref{eq:Tla-xi}) and (\ref{eq:heatflux-xi-h0}), Eq. (\ref{eq:geq-Hl}) is solved. Finally, substituting the obtained solution $H_{l}$ into Eqs. (\ref{eq:amplificationrate}) and (\ref{eq:phasevelocity}) and replacing $\mu_{l}$ with $(h_{0}/\delta_{0})\mu_{a}$, it is possible to calculate the amplification rate $\sigma_{*}^{(r)}$ and phase velocity $v_{p*}$ with respect to $\mu_{a}$. 

\section{Results}

\subsection{\label{sec:solutionTa}Solutions of temperature distributions in the air boundary layer}

\begin{figure}[ht]
\begin{center}
\includegraphics[width=7cm,height=7cm,keepaspectratio,clip]{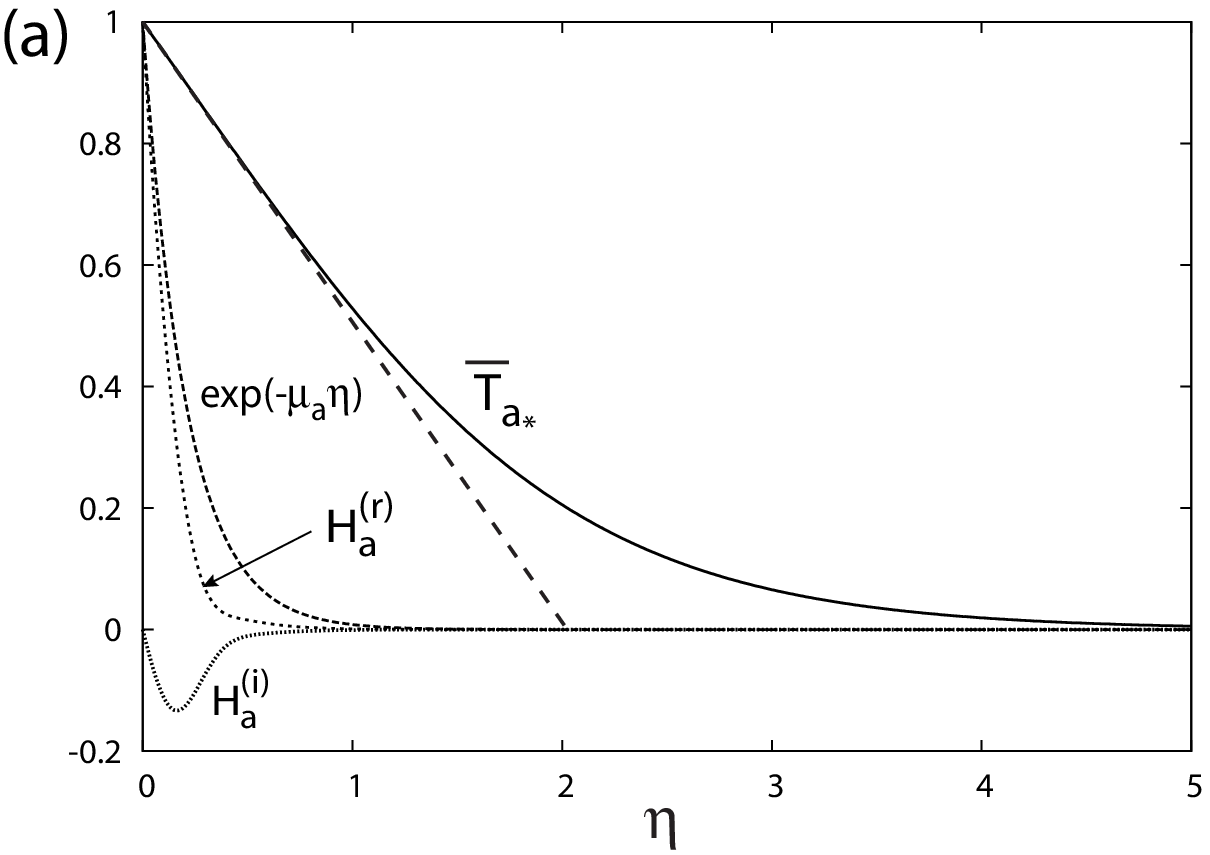}\qquad
\includegraphics[width=7cm,height=7cm,keepaspectratio,clip]{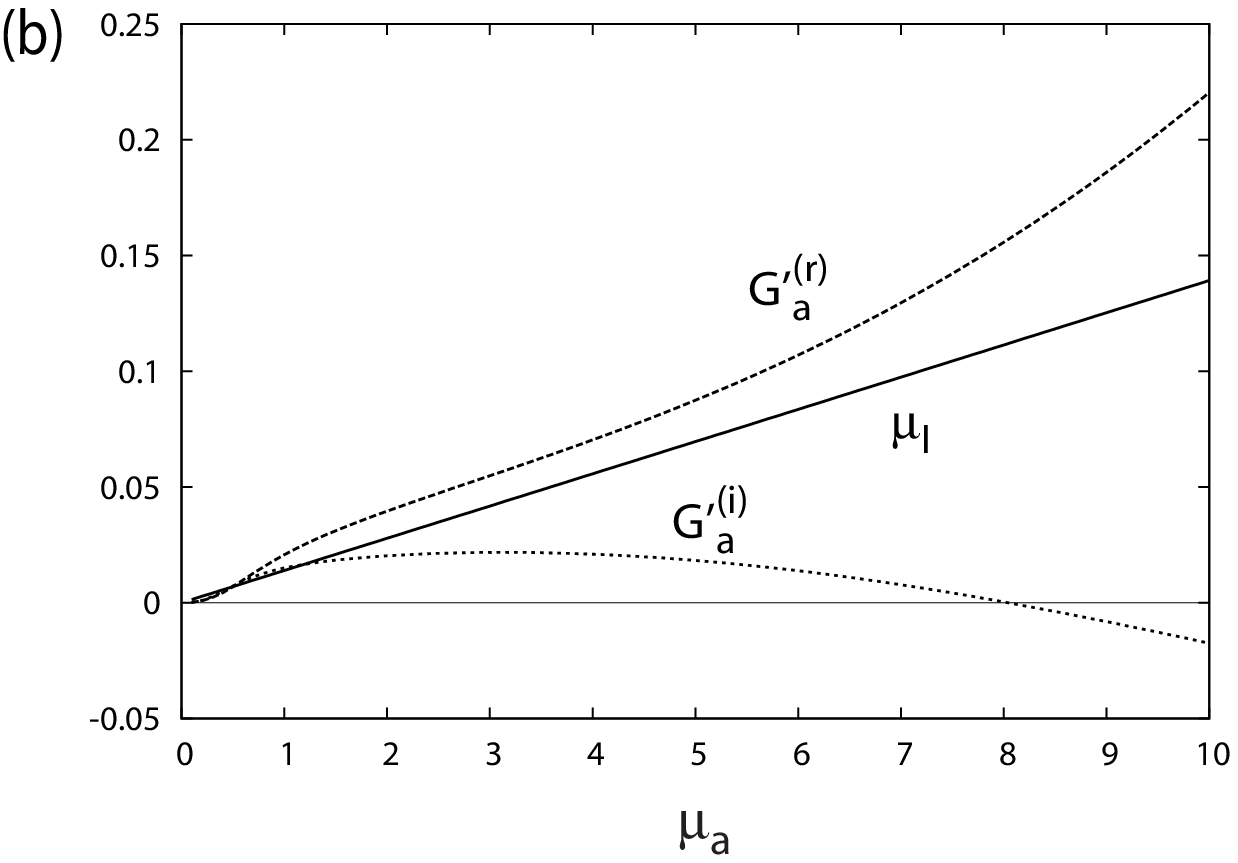}
\end{center}
\caption{For $Q/l=50$ $[{\rm (ml/h)/cm}]$, $\theta=\pi/2$, $x=1.0$ m and $\Delta T_{a}=10$ $^{\circ}$C,
(a) air temperature distribution $\bar{T}_{a*}$, and distributions of ${\rm exp}(-\mu_{a}\eta)$, $H_{a}^{(r)}$ and $H_{a}^{(i)}$ at the dimensionless wave number of $\mu_{a}=4.8$.
(b) perturbed part of air temperature gradient $G'_{a}\equiv h_{0}/\delta_{0}(-dH_{a}/d\eta|_{\eta=0})$ at the water-air surface: 
in the absence of airflow $G'_{a}=\mu_{l}$; 
in the presence of airflow $G'^{(r)}_{a}$ and $G'^{(i)}_{a}$ are the real and imaginary parts of $G'_{a}$. Here $\mu_{a}=10$ corresponds to the wavelength of 4.1 mm when $\delta_{0}=6.6$ mm.}
\label{fig:tempprofiles-Ga}
\end{figure}

In the absence of airflow, Eq. (\ref{eq:geq-Ha}) yields $d^{2}H_{a}/d\eta^{2}=\mu_{a}^{2}H_{a}$. With the boundary conditions $H_{a}|_{\eta=0}=1$ and $H_{a}|_{\eta=\infty}=0$, the solution is given by $H_{a}={\rm exp}(-\mu_{a}\eta)$, and hence 
$G'^{(r)}_{a}=(h_{0}/\delta_{0})\mu_{a}=kh_{0}=\mu_{l}$ and $G'^{(i)}_{a}=0$. 
In the presence of airflow, as shown in Fig. \ref{fig:tempprofiles-Ga} (a), $H_{a}^{(r)}$ decreases more rapidly than the exponential function, and $H_{a}^{(i)}$ acquires non-zero values. Therefore, as shown in Fig. \ref{fig:tempprofiles-Ga} (b), the value of $G'^{(r)}_{a}$ is greater than $\mu_{l}$ and $G'^{(i)}_{a}$ acquires non-zero values. 

From Eq. (\ref{eq:heatflux-xi}), the energy conservation equation at the unperturbed water-air surface is $-K_{l}\partial\bar{T}_{l}/\partial y|_{y=h_{0}}=-K_{a}\partial\bar{T}_{a}/\partial y|_{y=h_{0}}$. When the linear temperature profile $\bar{T}_{l}$ in the water layer and the exact temperature profile $\bar{T}_{a*}$ in the air boundary layer are substituted into the above energy conservation equation,
$T_{la}$ in Eq. (\ref{eq:Tla}) is obtained as 
\begin{equation}
T_{la} \approx T_{sl}+\frac{K_{a}}{K_{l}}\frac{h_{0}}{\delta_{0}/\bar{G}_{a*}}T_{\infty}.
\label{eq:Tla-airflow}
\end{equation}
From Eq. (\ref{eq:heatflux-zeta}), the energy conservation equation at the unperturbed ice-water interface is  
$L\bar{V}=K_{l}(T_{sl}-T_{la})/h_{0}$. 
Substituting  Eq. (\ref{eq:Tla-airflow}) into this equation 
yields 
\begin{equation}
\bar{V} \approx -\frac{K_{a}T_{\infty}}{L(\delta_{0}/\bar{G}_{a*})}.
\label{eq:V-airflow}
\end{equation}
If $\delta_{0}/\bar{G}_{a*}$ in Eqs. (\ref{eq:Tla-airflow}) and (\ref{eq:V-airflow}) is considered as $\delta$ represented in Fig. \ref{fig:ice-water-air} (a), then $T_{la}$ and $\bar{V}$ in the previous papers \cite{Ueno04, Ueno07, Ueno09, Short06} or $\bar{V}$ mentioned in the Introduction in this paper are obtained. 
The linear temperature profile in the air assumed in the previous papers \cite{Ueno04, Ueno07, Ueno09, Short06}
is shown by the dashed line in Fig. \ref{fig:tempprofiles-Ga} (a), which is expressed as $\bar{T}_{a*}=1-\bar{G}_{a*}\eta$. Here, $\bar{G}_{a*}=-d\bar{T}_{a*}/d\eta|_{\eta=0} $ can be estimated numerically yielding a value of about 0.5.
Using our notation, the boundary layer thickness in paper \cite{Short06} is expressed as $\delta=C\delta_{0}/\sqrt{2}$, 
which must be equal to $\delta=\delta_{0}/\bar{G}_{a*}$.
From this, the parameter $C$ is determined as  $C=\sqrt{2}/\bar{G}_{a*}\approx 2.8$. 
Since $\bar{G}_{a*}$ is obtained from the solution of Eqs. (\ref{eq:geq-basicFa}) and  (\ref{eq:geq-basicTa}), $\delta$ depends on the Prandtl number of air.  
 
\subsection{Approximate solutions of flow and temperature distributions in the water layer}

Since $\delta_{0}$ is of the same order as the characteristic length scale of ripples, we cannot use the long wavelength approximation, the higher order of $\mu_{a}$ in Eqs. (\ref{eq:geq-fa}) and (\ref{eq:geq-Ha}) have to be retained. On the other hand, since the water layer thickness $h_{0}$ is much less than the characteristic length scale of ripples, we can neglect the higher order of $\mu_{l}$ in Eqs. (\ref{eq:geq-fl}), (\ref{eq:geq-Hl}), (\ref{eq:shear-stress-xi}) and (\ref{eq:normal-stress-xi}). Using the long wavelength approximation, $f_{l}$ and $H_{l}$ can be calculated approximately as in the previous papers. \cite{Ueno03, Ueno04, Ueno07, Ueno09}
 
Transferring the variable $y_{*}$ to $z=1-y_{*}$, the general solution of (\ref{eq:geq-Hl}) is expressed as: \cite{Ueno03, Ueno09}
\begin{equation}
H_{l}(z)=C_{1}\phi_{1}(z)+C_{2}\phi_{2}(z)+i\mu_{l} \Pec_{l}\int_{0}^{z}\left\{\phi_{2}(z)\phi_{1}(z')-\phi_{1}(z)\phi_{2}(z')\right\}f_{l}(z')dz', 
\label{eq:sol-Hl}  
\end{equation}
where $\phi_{1}$ and $\phi_{2}$ are solutions of the homogeneous equation (\ref{eq:geq-Hl}). 
From Eqs. (\ref{eq:Tla-xi}) and (\ref{eq:heatflux-xi-h0}), we obtain 
$C_{1}=f_{l}|_{z=0}$ and
$C_{2}=h_{0}/\delta_{0}(-dH_{a}/d\eta|_{\eta=0})f_{l}|_{z=0}$, respectively,
because $\phi_{1}|_{z=0}=1$, $\phi_{2}|_{z=0}=0$, $d\phi_{1}/dz|_{z=0}=0$ and $d\phi_{2}/dz|_{z=0}=1$.
Consequently, $H_{l}$ is expressed as
\begin{eqnarray}
H_{l}(z)
&=&f_{l}|_{z=0}\left\{\phi_{1}(z)+\frac{h_{0}}{\delta_{0}}\left(-\frac{dH_{a}}{d\eta}\Big|_{\eta=0}\right)\phi_{2}(z)\right\} 
\nonumber \\
&& +i\mu_{l} \Pec_{l} \int_{0}^{z}\left\{\phi_{2}(z)\phi_{1}(z')-\phi_{1}(z)\phi_{2}(z')\right\}f_{l}(z')dz'. 
\label{eq:finalsol-Hl}  
\end{eqnarray}

For typical values of $h_{0}$ and $u_{l0}$, $\Rey_{l}\sim 1$ and $\Pec_{l} \sim 10$; then $\mu_{l}\Rey_{l} \ll 1$ and $\mu_{l}\Pec_{l} \sim 1$ for the length scale of ripples on icicles. Therefore, we can neglect the $\mu_{l}\Rey_{l}$ term in Eqs. (\ref{eq:geq-fl}) and (\ref{eq:normal-stress-xi}). This corresponds to neglecting the inertia term of the full Orr-Sommerfeld equation. \cite{Ueno07} Furthermore, the expansion of $\phi_{1}$ and $\phi_{2}$ with respect to $\mu_{l}\Pec_{l}$ 
up to the first order is sufficient. Indeed, the justification for these approximations was confirmed by our recent numerical analysis. \cite{Ueno09} Hence, it is sufficient to use the following approximate solutions:
\begin{equation}
f_{l}(z)=\frac{1}{6+i\alpha}(6+i\alpha z-6z^{2}-i\alpha z^{3}),
\label{eq:fl}
\end{equation}
\begin{equation}
\phi_{1}(z) 
=1-i\left(\frac{1}{2}z^{2}-\frac{1}{12}z^{4}\right)\mu_{l}\Pec_{l}, 
\label{eq:phi1}
\end{equation}
\begin{equation}
\phi_{2}(z) 
=z-i\left(\frac{1}{6}z^{3}-\frac{1}{20}z^{5}\right)\mu_{l} \Pec_{l}.
\label{eq:phi2}
\end{equation}
Since the direction of the $x$ axis in Fig. \ref{fig:ice-water-air} is opposite to that in the previous papers, \cite{Ueno03, Ueno04, Ueno07, Ueno09} we note that the sign of $\bar{U}_{l*}$ in this paper is opposite. This leads to different functional forms of $f_{l}$, $\phi_{1}$ and $\phi_{2}$ from those in the previous papers. 

In the presence of airflow, using the approximate solutions (\ref{eq:fl}), (\ref{eq:phi1}) and (\ref{eq:phi2}), Eqs. (\ref{eq:amplificationrate}) and (\ref{eq:phasevelocity}) yield  
\begin{eqnarray}
\sigma_{*}^{(r)}&=& 
\frac{
G'^{(r)}_{a}\left\{36-\frac{3}{2}\alpha(\mu_{l}\Pec_{l})\right\}
+G'^{(i)}_{a}\left\{6\alpha+9\mu_{l}\Pec_{l}\right\}
-\frac{3}{2}\alpha(\mu_{l}\Pec_{l})}{36+\alpha^{2}}\nonumber \\ 
&&+K^{s}_{l}\mu_{l}
\frac{
G'^{(r)}_{a}\left\{36-\frac{7}{10}\alpha(\mu_{l}\Pec_{l})\right\}
+G'^{(i)}_{a}\left\{6\alpha+\frac{21}{5}\mu_{l}\Pec_{l}\right\}
-\frac{7}{10}\alpha(\mu_{l}\Pec_{l})-\alpha^{2}}
{36+\alpha^{2}},
\label{eq:amp-airflow}
\end{eqnarray}
\begin{eqnarray}
v_{p*}&=&\frac{1}{\mu_{l}}\left[
\frac{-\frac{1}{4}\alpha^{2}(\mu_{l}\Pec_{l})
+G'^{(r)}_{a}\left\{6\alpha+9\mu_{l}\Pec_{l}\right\}
-G'^{(i)}_{a}\left\{36-\frac{3}{2}\alpha(\mu_{l}\Pec_{l})\right\}}{36+\alpha^{2}} \right. \nonumber \\
&& \left.+K^{s}_{l}\mu_{l}
\frac{6\alpha-\frac{7}{60}\alpha^{2}(\mu_{l}\Pec_{l})
+G'^{(r)}_{a}\left\{6\alpha+\frac{21}{5}\mu_{l}\Pec_{l}\right\}
-G'^{(i)}_{a}\left\{36-\frac{7}{10}\alpha(\mu_{l}\Pec_{l})\right\}}
{36+\alpha^{2}}\right]. 
\label{eq:vp-airflow}
\end{eqnarray}
On the other hand, in the absence of airflow, since $G'^{(r)}_{a}=\mu_{l}$ and $G'^{(i)}_{a}=0$ as mentioned above, Eqs. (\ref{eq:amp-airflow}) and (\ref{eq:vp-airflow}) reduce to the previous dispersion relation: \cite{Ueno03, Ueno09}
\begin{eqnarray}
\sigma_{*}^{(r)}
&=& 
\frac{
\mu_{l}\left\{36-\frac{3}{2}\alpha(\mu_{l}\Pec_{l})\right\}
-\frac{3}{2}\alpha(\mu_{l}\Pec_{l})}{36+\alpha^{2}} 
+K^{s}_{l}\mu_{l}
\frac{
\mu_{l}\left\{36-\frac{7}{10}\alpha(\mu_{l}\Pec_{l})\right\}
-\frac{7}{10}\alpha(\mu_{l}\Pec_{l})-\alpha^{2}}{36+\alpha^{2}},\nonumber \\
\label{eq:Uamp-noairflow}
\end{eqnarray}
\begin{eqnarray}
v_{p*}&=&\frac{1}{\mu_{l}}
\left[\frac{-\frac{1}{4}\alpha^{2}(\mu_{l}\Pec_{l})
+\mu_{l}\left\{6\alpha+9\mu_{l}\Pec_{l}\right\}}{36+\alpha^{2}}  
+K^{s}_{l}\mu_{l}\frac{6\alpha-\frac{7}{60}\alpha^{2}(\mu_{l}\Pec_{l})
+\mu_{l}\left\{6\alpha+\frac{21}{5}\mu_{l}\Pec_{l}\right\}}{36+\alpha^{2}}\right].\nonumber \\
\label{eq:Uvp-noairflow}
\end{eqnarray}

\subsection{\label{sec:wavelength}Wavelength and translation velocity of ripples}

\begin{figure}[ht]
\begin{center}
\includegraphics[width=7cm,height=7cm,keepaspectratio,clip]{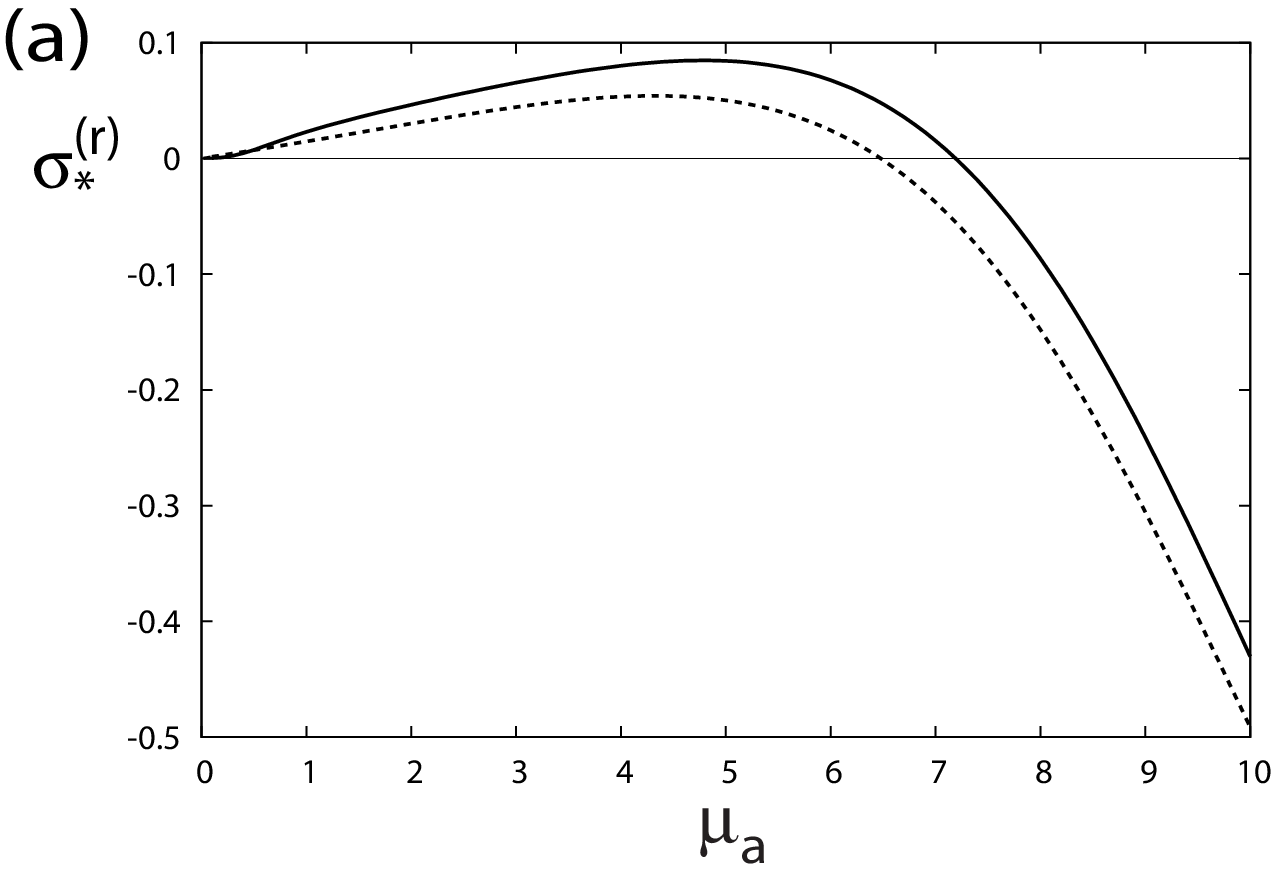}
\hspace{1cm}
\includegraphics[width=7cm,height=7cm,keepaspectratio,clip]{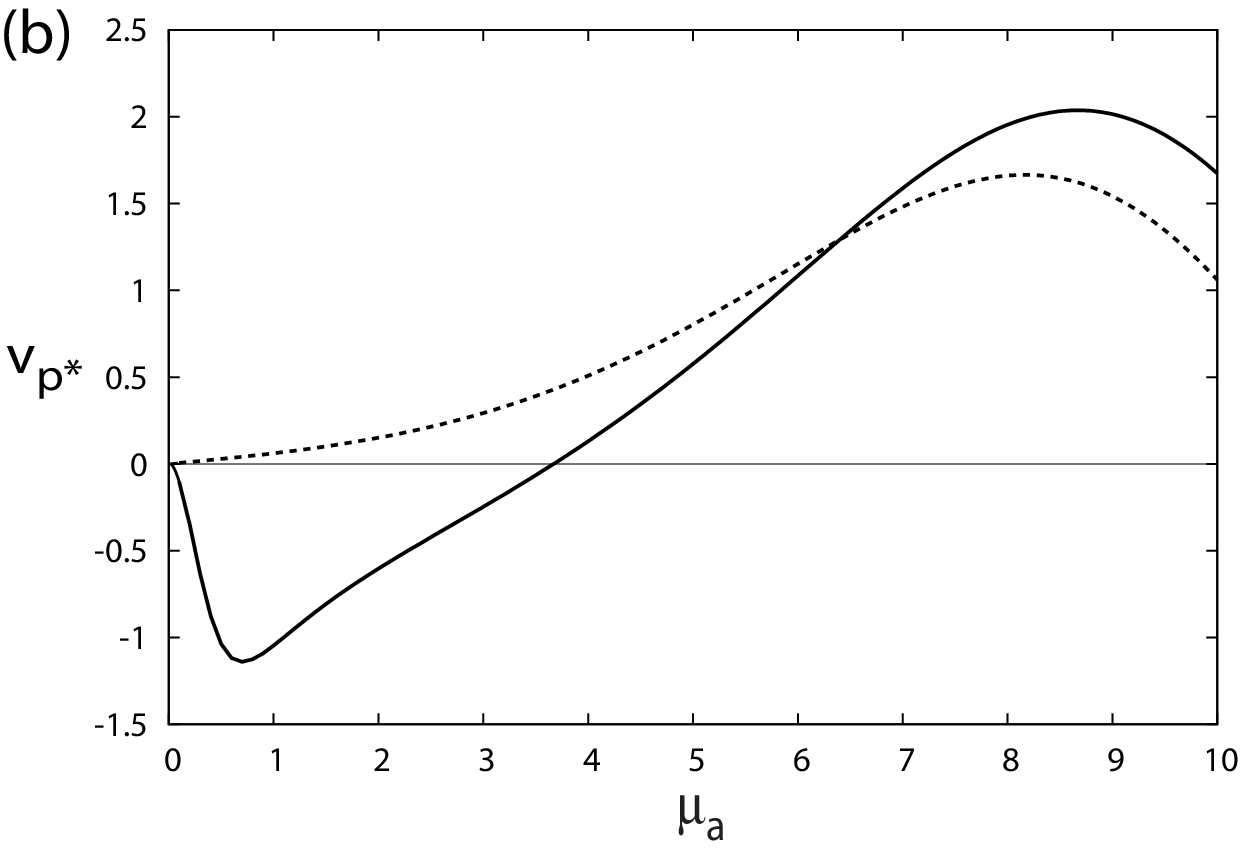}
\\[5mm] 
\includegraphics[width=8cm,height=8cm,keepaspectratio,clip]{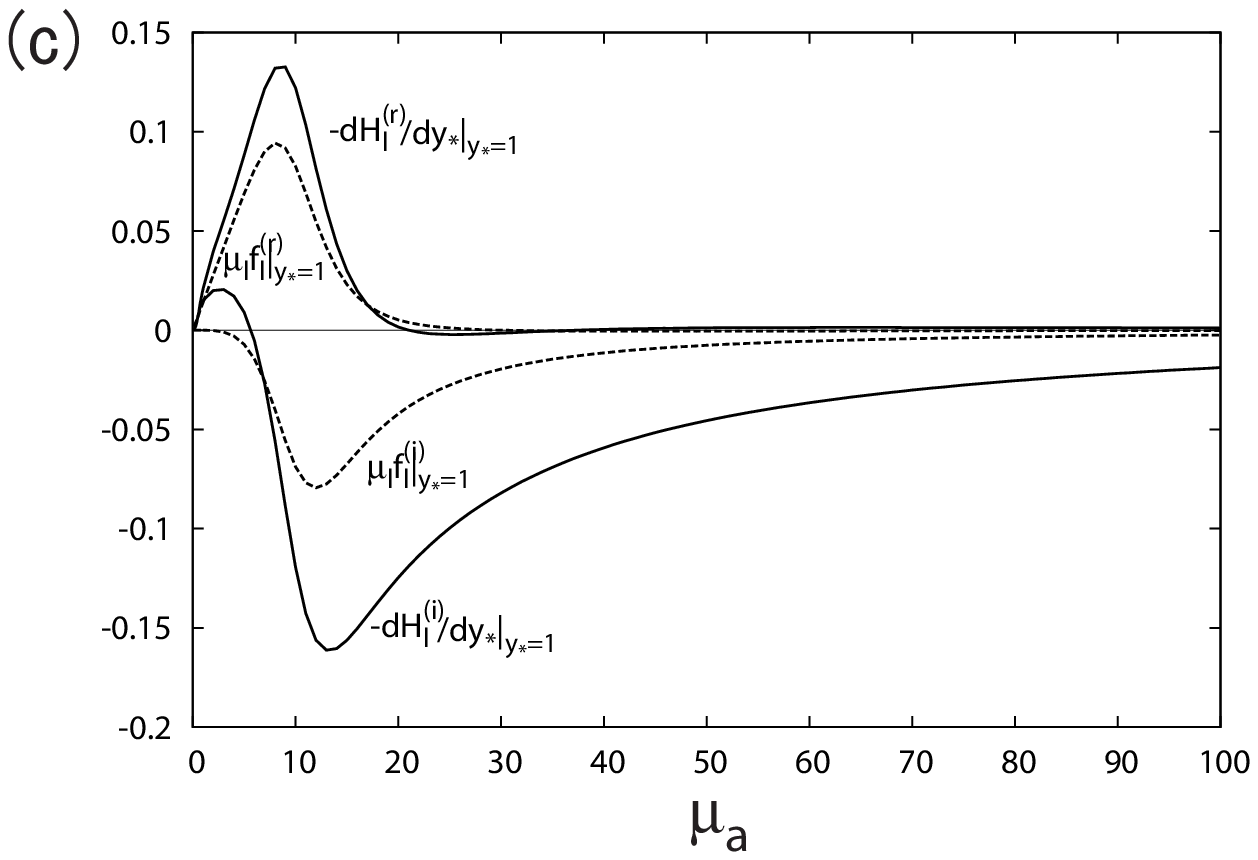}
\end{center}
\caption{For $Q/l=50$ $[{\rm (ml/h)/cm}]$, $\theta=\pi/2$ and $\delta_{0}=6.6$ mm, 
(a) dimensionless amplification rate $\sigma_{*}^{(r)}=\sigma^{(r)}/(\bar{V}/h_{0})$ versus dimensionless  wave number $\mu_{a}=k\delta_{0}$; 
(b) dimensionless  phase velocity $v_{p*}=v_{p}/\bar{V}$ versus dimensionless  wave number $\mu_{a}$. Solid and dashed lines indicate the presence and absence of airflow, respectively.
(c) The behaviour of the real and imaginary parts of the perturbed temperature gradient at the water-air surface with respect to $\mu_{a}$, in the presence of airflow (solid lines) and in the absence of airflow (dashed lines).
Here $\mu_{a}=100$ corresponds to the wavelength of 413 $\mu$m when $\delta_{0}=6.6$ mm.}
\label{fig:sim-mua-amp-vp}
\end{figure}

For the water supply rate per width $Q/l=50$ [(ml/h)/cm] and the angle $\theta=\pi/2$, Figs. \ref{fig:sim-mua-amp-vp} (a) and (b) show numerically obtained the dimensionless amplification rate $\sigma_{*}^{(r)}=\sigma^{(r)}/(\bar{V}/h_{0})$ and the dimensionless translation velocity $v_{p*}=v_{p}/\bar{V}$ versus dimensionless wave number $\mu_{a}=k\delta_{0}$, respectively. The wave number of ripples that one expects to observe is that for which the amplification rate is the maximum. We also define the value of $v_{p*}$ from the wave number at which $\sigma_{*}^{(r)}$ acquires a maximum value. 
In the presence of airflow, $\sigma_{*}^{(r)}$ acquires a maximum value of $\sigma^{(r)}_{*\rm max}=0.085$ at $\mu_{a}=4.8$ (solid line in Fig. \ref{fig:sim-mua-amp-vp} (a)). Since the wave number $k$ is normalized by $\delta_{0}$, the corresponding wavelength is 8.6 mm from $\lambda=2\pi\delta_{0}/\mu_{a}$. Here we have used $\delta_{0}=6.6$ mm estimated from the two parameters $x=1.0$ m and $\Delta T_{a}=10$ $^{\circ}$C. 
At $\mu_{a}=4.8$, $v_{p*}=0.48$ as represented by the solid line in Fig. \ref{fig:sim-mua-amp-vp} (b). 
On the other hand, in the absence of airflow, Eq. (\ref{eq:Uamp-noairflow}) acquires a maximum value of $\sigma^{(r)}_{*\rm max}=0.054$ at $\mu_{a}=4.3$ (dashed line in Fig. \ref{fig:sim-mua-amp-vp} (a)), which corresponds to the wavelength of $\lambda=9.6$mm. At $\mu_{a}=4.3$, $v_{p*}=0.59$ as represented by the dashed line in Fig. \ref{fig:sim-mua-amp-vp} (b).

Any disturbance near the solidification front can be initiated by non-uniformity in temperature in the vicinity of the ice-water interface. Since the water layer considered here is very thin, we cannot neglect the influence of external disturbance at the water-air surface on the growth condition of the ice-water interface. 
In order to determine the growth condition from the dispersion relation (\ref{eq:dispersion}), it is necessary to obtain the perturbed temperature amplitude $H_{l}$ in the water layer. $H_{l}$ must satisfy the boundary condition (\ref{eq:heatflux-xi-h0}) which includes the perturbed air temperature gradient at the water-air surface.
Using Eq. (\ref{eq:Gar-Gai}) and $\bar{U}_{l*}|_{y_{*}=1}=-1$, the real and imaginary parts of Eq. (\ref{eq:heatflux-xi-h0}) can be written as follows:
\begin{equation}
-\frac{dH^{(r)}_{l}}{dy_{*}}\Big|_{y_{*}=1}=G'^{(r)}_{a}f^{(r)}_{l}|_{y_{*}=1}-G'^{(i)}_{a}f^{(i)}_{l}|_{y_{*}=1}, \qquad
-\frac{dH^{(i)}_{l}}{dy_{*}}\Big|_{y_{*}=1}=G'^{(r)}_{a}f^{(i)}_{l}|_{y_{*}=1}+G'^{(i)}_{a}f^{(r)}_{l}|_{y_{*}=1}.
\label{eq:dHlr-dHli-airflow}
\end{equation}
Since $G'^{(r)}_{a}=\mu_{l}$ and $G'^{(i)}_{a}=0$ in the absence of airflow, Eq. (\ref{eq:dHlr-dHli-airflow}) reduces to the previous results: \cite{Ueno09}
\begin{equation}
-\frac{dH^{(r)}_{l}}{dy_{*}}\Big|_{y_{*}=1}=\mu_{l}f^{(r)}_{l}|_{y_{*}=1}, \qquad
-\frac{dH^{(i)}_{l}}{dy_{*}}\Big|_{y_{*}=1}=\mu_{l}f^{(i)}_{l}|_{y_{*}=1}. 
\label{eq:dHlr-dHli-noairflow}
\end{equation}
The solid and dashed lines in Fig. \ref{fig:sim-mua-amp-vp} (c) show the behaviour of 
$-dH^{(r)}_{l}/dy_{*}|_{y_{*}=1}$, $-dH^{(i)}_{l}/dy_{*}|_{y_{*}=1}$ in Eq. (\ref{eq:dHlr-dHli-airflow}) 
and of 
$\mu_{l}f^{(r)}_{l}|_{y_{*}=1}$, $\mu_{l}f^{(i)}_{l}|_{y_{*}=1}$ in Eq. (\ref{eq:dHlr-dHli-noairflow}) 
with respect to $\mu_{a}$. It can be seen that $-dH^{(r)}_{l}/dy_{*}|_{y_{*}=1}$ and $\mu_{l}f^{(r)}_{l}|_{y_{*}=1}$ increase for small $\mu_{a}$. 
In the absence of airflow, the rate of latent heat loss due to thermal diffusion from the water-air surface to the air changes locally by the water-air surface disturbance. \cite{Ueno09} On the other hand, in the presence of airflow, the rate of latent heat loss is enhanced by the airflow, more so than in the case of thermal diffusion. However, as shown in Fig. \ref{fig:sim-mua-amp-vp} (c), non-uniformity of the rate of latent heat loss at the water-air surface decreases with an increase in $\mu_{a}$ because of the action of the restoring force on the water-air surface, which causes the amplitude of the water-air surface disturbance to decrease. \cite{Ueno03, Ueno04, Ueno07, Ueno09} This effect is due to the parameter $\alpha$ in $f_{l}$ in Eqs. (\ref{eq:dHlr-dHli-airflow}) and (\ref{eq:dHlr-dHli-noairflow}) and is more effective for large wave numbers.
The physical meaning that the values of $-dH^{(i)}_{l}/dy_{*}|_{y_{*}=1}$ in Eq. (\ref{eq:dHlr-dHli-airflow}) and $\mu_{l}f^{(i)}_{l}|_{y_{*}=1}$ in Eq. (\ref{eq:dHlr-dHli-noairflow}) are not zero will be discussed in \ref{sec:heatflux}.

An approximation of Eq. (\ref{eq:amp-airflow}) makes the above discussion more clear.
We note that the second term in Eq. (\ref{eq:amp-airflow}) is smaller than the first term, and the wave number at which $\sigma_{*}^{(r)}$ acquires a maximum value is almost the same as that without the second term. \cite{Ueno09}
Therefore, extracting the most dominant term from the first term in (\ref{eq:amp-airflow}) and using (\ref{eq:alpha}), we obtain 
\begin{equation}
\sigma_{*}^{(r)}
\approx 
\frac{36G'^{(r)}_{a}-\frac{3}{2}\alpha(\mu_{l}\Pec_{l})}{36}=G'^{(r)}_{a}-\frac{\Pec_{l}}{12}\left(\frac{a}{h_{0}}\right)^{2}\mu_{l}^{4},
\label{eq:amp-approx-airflow}
\end{equation}
at $\theta=\pi/2$. As mentioned above, the non-uniformity of the air temperature gradient at the water-air surface is the trigger of the ice-water interface instability, which is represented by the positive term $G'^{(r)}_{a}$ in Eq. (\ref{eq:amp-approx-airflow}).
In the absence of airflow, since $G'^{(r)}_{a}=\mu_{l}$, we find from $d\sigma_{*}^{(r)}/d\mu_{l}=0$ that $\sigma_{*}^{(r)}$ acquires a maximum value at $\mu_{l}=[3(h_{0}/a)^2/\Pec_{l}]^{1/3}$. From this, an approximate formula is obtained to determine the wavelength of the ripples: $\lambda=2\pi(a^{2}h_{0}\Pec_{l}/3)^{1/3}$, \cite{Ueno07, Ueno09} as mentioned in the Introduction in this paper. 
On the other hand, in the presence of airflow, the value of $G'^{(r)}_{a}$ is greater than $\mu_{l}$, as shown in Fig. \ref{fig:tempprofiles-Ga} (b). This indicates that the natural convection airflow enhances the destabilization of the ice-water interface compared to the destabilization due to the thermal diffusion. However, it is difficult to express the dependence of $G'^{(r)}_{a}$ on $\mu_{a}$ analytically. 
The stabilization of the ice-water interface is dominated by the negative term in Eq. (\ref{eq:amp-approx-airflow}). The stabilization mechanism due to the action of the restoring force of the surface tension and gravity on the water-air surface is not relevant to the airflow. Although the value of $\sigma^{(r)}_{*\rm max}$ in the presence of airflow is greater than that in its absence, the wavelengths determined from the most unstable mode have nearly the same value in both cases.
However, there is a considerable difference in $v_{p*}$. In the absence of airflow, $v_{p*}>0$ for all $\mu_{a}$, as shown by the dashed line in Fig. \ref{fig:sim-mua-amp-vp} (b). On the other hand, in the presence of airflow, $v_{p*}$ has negative values for a small wave number region because the terms with $G'^{(i)}_{a}$ in Eq. (\ref{eq:vp-airflow}) are the most dominant. The solid line in Fig. \ref{fig:sim-mua-amp-vp} (b) indicates that the sign of $v_{p*}$ changes from negative to positive at $\mu_{a}=3.7$. What determines the sign of $v_{p*}$ will be discussed in \ref{sec:heatflux}.   

\begin{figure}[ht]
\begin{center}
\includegraphics[width=7cm,height=7cm,keepaspectratio,clip]{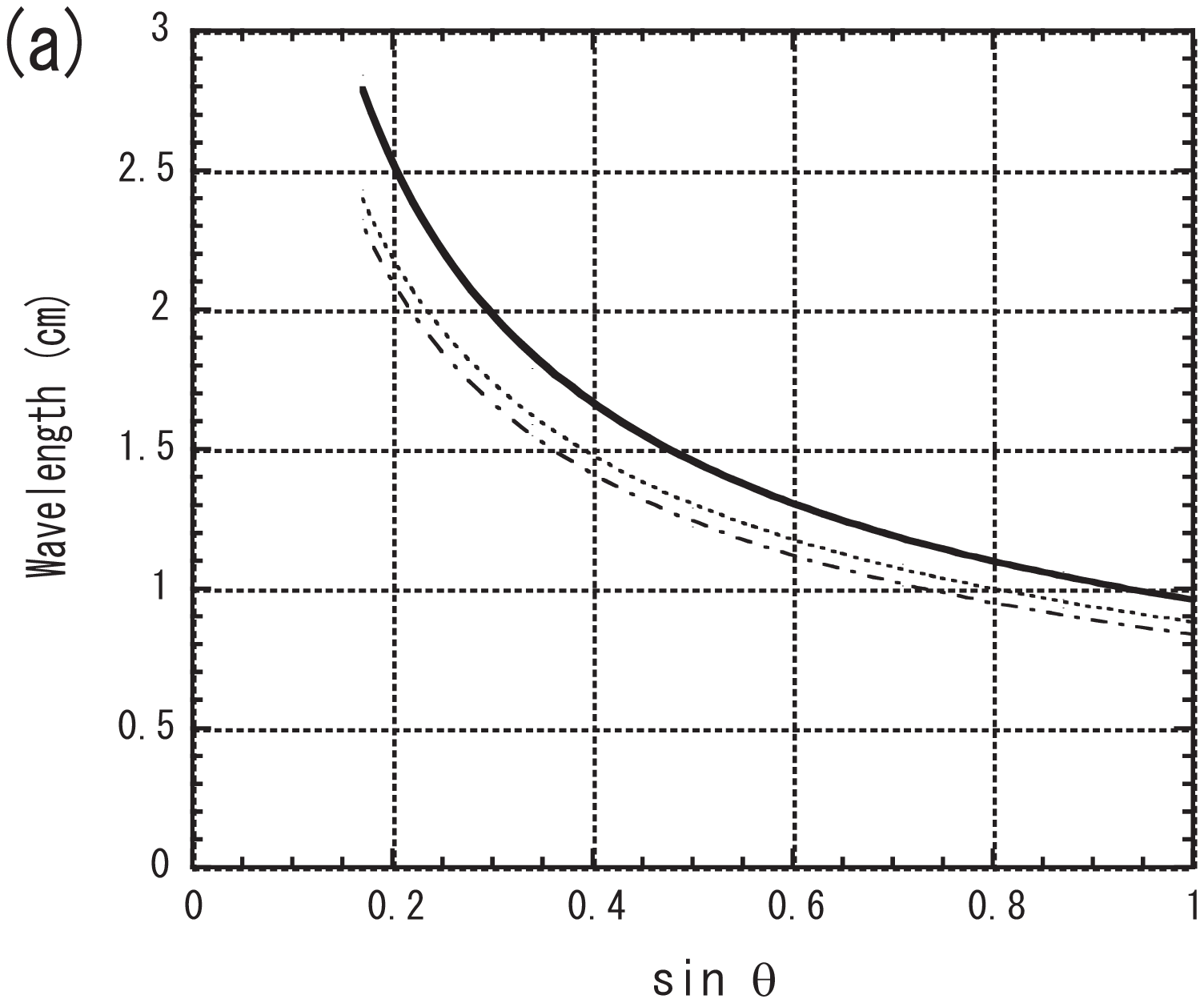}
\hspace{2mm}
\includegraphics[width=7cm,height=7cm,keepaspectratio,clip]{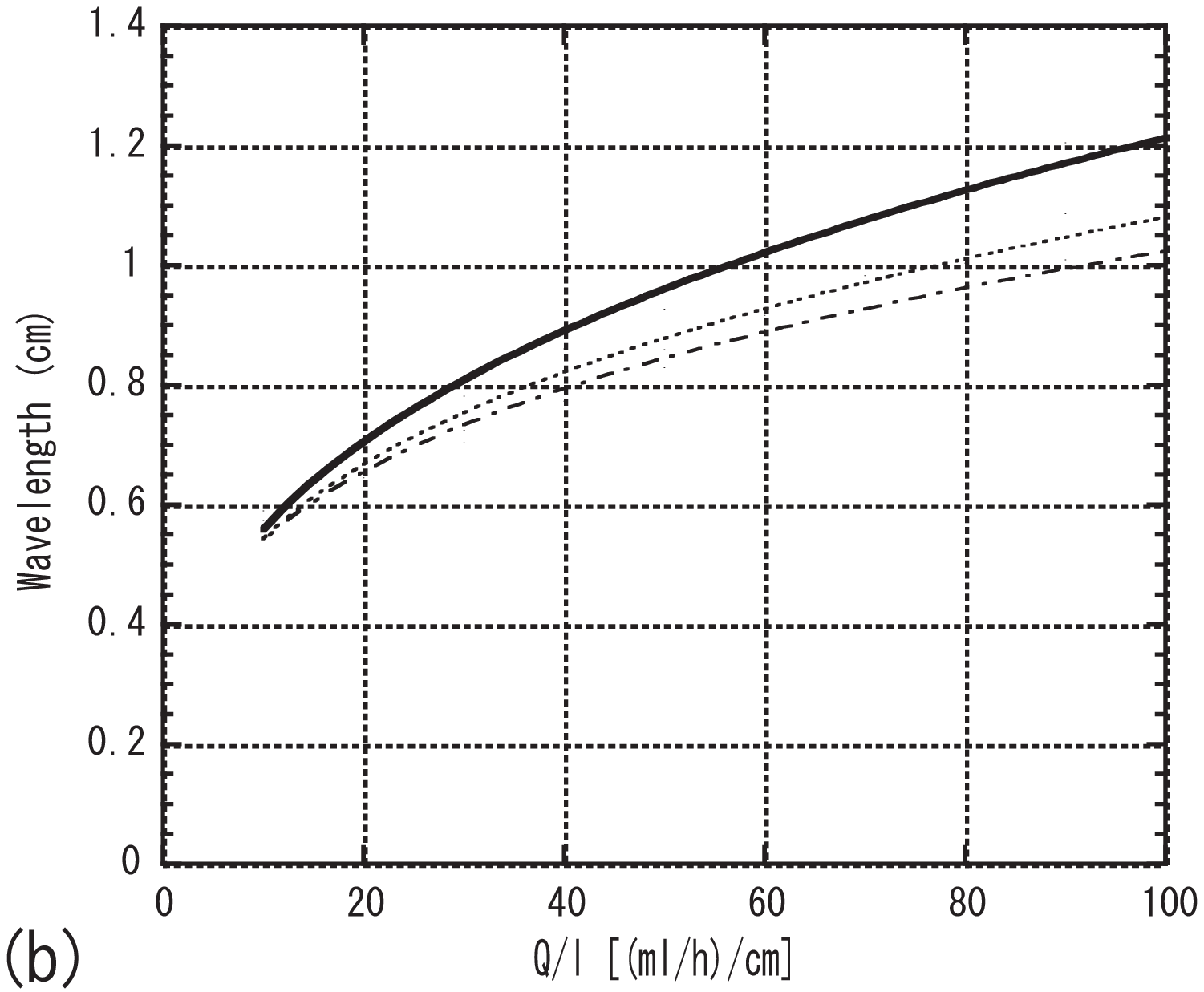}
\\[5mm]
\includegraphics[width=7cm,height=7cm,keepaspectratio,clip]{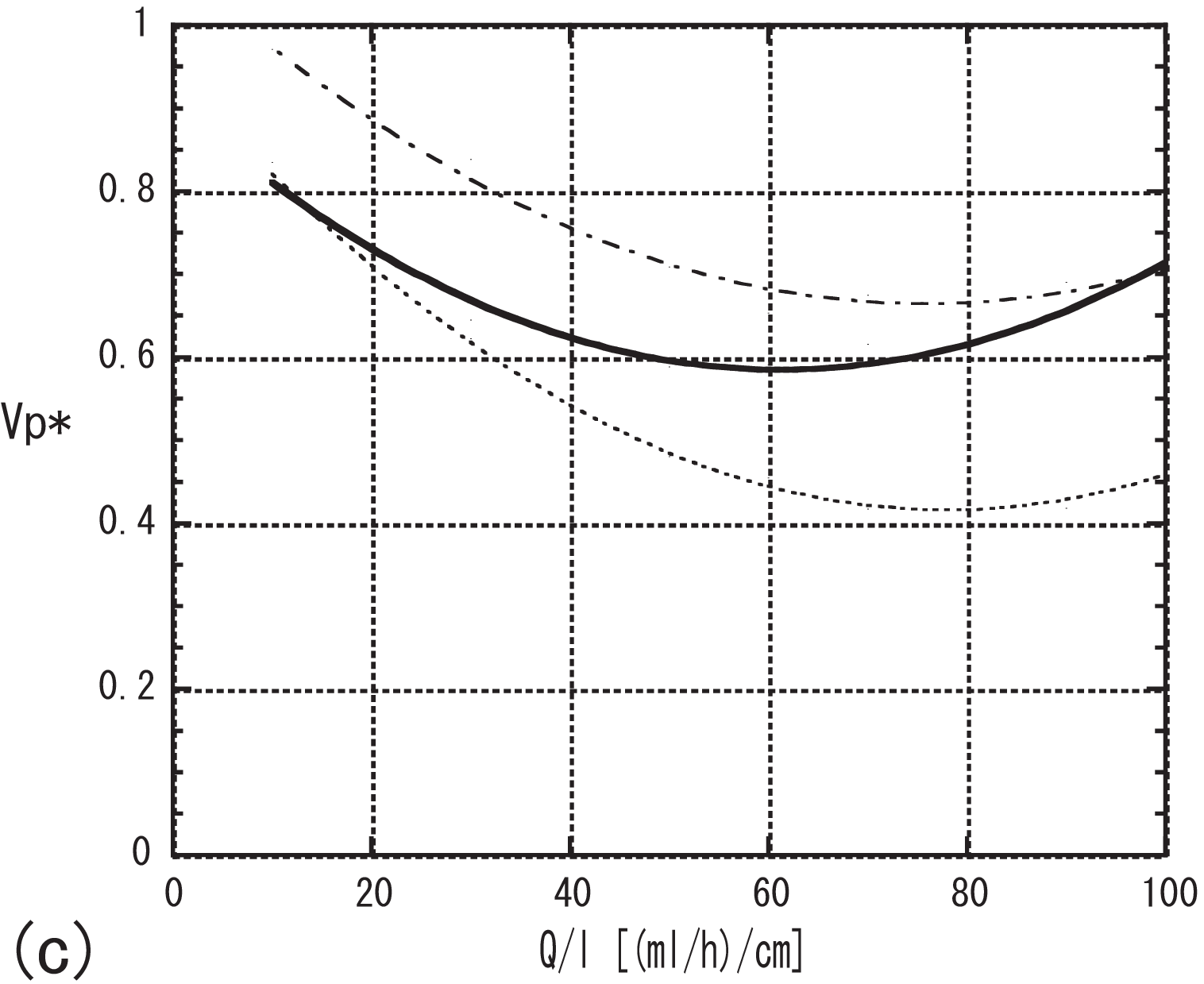}
\end{center}
\caption{(a) The wavelength versus $\sin\theta$ at $Q/l=$160/3 [(ml/h)/cm]. 
(b) The wavelength versus $Q/l$ at $\theta=\pi/2$. 
(c) The phase velocity versus $Q/l$ at $\theta=\pi/2$.
Solid lines indicate the absence of airflow; \cite{Ueno03, Ueno07}
dashed lines and dashed-dotted lines indicate the presence of airflow for $Gr=609$ and $Gr=108$.}
\label{fig:wavelength-angle-Qoverl-airflow}
\end{figure}

Figures \ref{fig:wavelength-angle-Qoverl-airflow} (a) and (b) show the dependence of the wavelength of ripples on $\sin\theta$ at $Q/l=160/3$ [(ml/h)/cm] and that on $Q/l$ [(ml/h)/cm] at $\theta=\pi/2$, respectively. We have determined these wavelengths from the value of $\mu_{a}$ at which $\sigma_{*}^{(r)}$ acquires a maximum value for a given $Q/l$ and $\theta$. The solid lines indicate the absence of airflow, whereas the dashed and dashed-dotted lines indicate the presence of airflow. 
Although the wavelengths of ripples in the presence of airflow are slightly shorter than those in the absence of airflow, the dependence of the wavelengths on the angles and water supply rates shows almost the same behaviour as the experimental results (see Figure 8  \cite{Ueno09}).
When determining the wavelengths in Fig. \ref{fig:wavelength-angle-Qoverl-airflow} (a) and (b), we have used 
$\delta_{0}=6.6$ mm ($x=1.0$ m and $\Delta T_{a}=10$ $^{\circ}$C) for the dashed lines and 
$\delta_{0}=3.7$ mm ($x=0.1$ m and $\Delta T_{a}=10$ $^{\circ}$C) for the dashed-dotted lines. 

Table \ref{tab:tableI} shows the wavelengths obtained from various values of $\delta_{0}=4x/Gr$ using different combination of $x$ and $\Delta T_{a}$. It is found that  the wavelength increases with the increase of both $x$ and $\Delta T_{a}$. This suggests that $G'^{(r)}_{a}$ in Eq. (\ref{eq:amp-approx-airflow}) must include the modified local Grashof number $Gr$. 
However, the dependence of the wavelength $\lambda$ on $x$ and $\Delta T_{a}$ in $Gr$ is extremely small compared to that of $\bar{V}$, $T_{la}$ and $\delta$. This result is relevant to the fact that the wavelength of ripples on icicles is nearly independent of the vertical position of icicles and ambient air temperature. 
Table \ref{tab:tableI} also shows that the value of $v_{p*}$ increases with a decrease in $\Delta T_{a}$ and a decrease in $x$.
Since $v_{p*}$ has positive values, the ripple with the most unstable mode moves only upwards. 
Figure \ref{fig:wavelength-angle-Qoverl-airflow} (c) shows the dependence of $v_{p*}$ on $Q/l$.  
The range of variation of $v_{p*}$ on $Q/l$ in the presence of airflow (dashed and dashed-dotted lines) is larger than that in the absence of airflow (solid line), and the dependence of $v_{p*}$ on $Gr$ is larger than that of the wavelength $\lambda$ on $Gr$. Therefore, we can say that $v_{p*}$ is sensitive to the parameters characterizing the air boundary layer.

\begin{table}[ht]
\caption{\label{tab:tableI} For $Q/l=50$ [(ml/h)/cm] and $\theta=\pi/2$, the dependence of modified local Grashof number, $Gr$, ice growth rate, $\bar{V}$, temperature at water-air surface, $T_{la}$, thickness of thermal boundary layer, $\delta=\delta_{0}/\bar{G}_{a*}$, wavelength of ripple, $\lambda$, and dimensionless translation velocity of ripple, $v_{p*}$, on air temperature far away, $T_{\infty}$, and position from the bottom of the gutter, $x$.}
\begin{ruledtabular}
\begin{tabular}{ccccccc}
&\multicolumn{4}{c}{$x=1.0$ m} \\
\hline
$T_{\infty}$ ($^{\circ}$C) & $Gr$ & $\bar{V}$ (mm/h) &  $T_{la}$ ($\times 10^{-3}$ $^{\circ}$C) 
& $\delta$ (mm) & $\lambda$ (mm) & $v_{p*}$ \\ 
\hline
 -5 & 512 & 0.08 & -1.2 & 16.1 & 8.3 & 0.65 \\
-10 & 609 & 0.20 & -2.9 & 13.4 & 8.6 & 0.48 \\
-15 & 674 & 0.32 & -4.9 & 12.1 & 8.7 & 0.41 \\
-20 & 724 & 0.47 & -7.0 & 11.2 & 8.7 & 0.37 \\
\hline
&\multicolumn{4}{c}{$x=0.1$ m} \\ 
\hline
 -5 &  91 & 0.14 &  -2.1 & 9.6 & 7.9 & 0.89 \\
-10 & 108 & 0.33 &  -5.0 & 7.9 & 8.3 & 0.71 \\
-15 & 120 & 0.56 &  -8.4 & 7.0 & 8.4 & 0.63 \\
-20 & 129 & 0.81 & -12.1 & 6.5 & 8.5 & 0.56 \\ 
\end{tabular}
\end{ruledtabular}
\end{table}

\subsection{\label{sec:heatflux}Heat flux at the ice-water interface and water-air surface}

We assume a dimensionless small perturbation of the ice-water interface with an infinitesimal initial amplitude $\delta_{b}=\zeta_{k}/h_{0}$: 
\begin{equation} 
y_{*}=\zeta_{*}=\delta_{b}\Imag[{\rm exp}(\sigma_{*}t_{*}+i\mu_{l}x_{*})]
=\delta_{b}(t_{*})\sin[\mu_{l}(x_{*}-v_{p*}t_{*})], 
\label{eq:zeta}
\end{equation}
where $\sigma_{*}=\sigma/(\bar{V}/h_{0})$, $t_{*}=(\bar{V}/h_{0})t$, $x_{*}=x/h_{0}$, $\delta_{b}(t_{*})\equiv \delta_{b}{\rm exp}(\sigma^{(r)}_{*}t_{*})$ and $\Imag$ denotes the imaginary part of its argument. 
The corresponding perturbation of the water-air surface with an infinitesimal initial amplitude $\delta_{t}=\xi_{k}/h_{0}$ is given by
\begin{eqnarray} 
y_{*}=\xi_{*}&=&1+\Imag[\delta_{t}{\rm exp}(\sigma_{*}t_{*}+i\mu_{l}x_{*})] \nonumber \\
&=&1+[(f_{l}^{(r)}|_{y_{*}=1})^{2}+(f_{l}^{(i)}|_{y_{*}=1})^2]^{1/2}
\delta_{b}(t_{*})\sin[\mu_{l}(x_{*}-v_{p*}t_{*})-\Theta_{\xi_{*}}],
\label{eq:xi}
\end{eqnarray}
where the relation $\delta_{t}=f_{l}|_{{y_{*}}=1}\delta_{b}$ for the amplitude is used, and $\Theta_{\xi_{*}}$ is a phase difference between the water-air surface and the ice-water interface.
Since $f_{l}|_{y_{*}=1}$ depends on the wave number through the parameter $\alpha$, the amplitude and phase of the water-air surface relative to the ice-water interface change depending on the wavelength of the ice-water interface disturbance. \cite{Ueno03, Ueno04, Ueno07, Ueno09}

The temperatures in the water layer and ice are expressed in the dimensionless forms: \cite{Ueno09}
\begin{equation}
T_{l*}(y_{*})\equiv \frac{T_{l}(y_{*})-T_{sl}}{T_{sl}-T_{la}}
=-y_{*}
+\delta_{b}\Imag[H_{l}(y_{*}){\rm exp}(\sigma_{*}t_{*}+i\mu_{l}x_{*})],
\label{eq:Tl}
\end{equation}
\begin{equation}
T_{s*}(y_{*}) \equiv \frac{T_{s}(y_{*})-T_{sl}}{T_{sl}-T_{la}}
=\delta_{b}{\rm exp}(\mu_{l} y_{*})\Imag[(H_{l}|_{y_{*}=0}-1){\rm exp}(\sigma_{*}t_{*}+i\mu_{l}x_{*})],
\label{eq:Ts}
\end{equation}
and the temperature in the air boundary layer is expressed as
\begin{equation}
T_{a*}(\eta)
=\bar{T}_{a*}(\eta)
+\Imag\left[\left(-\frac{d\bar{T}_{a*}}{d\eta}\Big|_{\eta=0}\right)
H_{a}(\eta)\frac{\xi_{k}}{\delta_{0}}{\rm exp}(\sigma_{*}t_{*}+i\mu_{l}x_{*})\right],
\label{eq:Ta}
\end{equation}
where we note that $y$ is normalized by $h_{0}$ in the water layer and ice, but $y$ is normalized by $\delta_{0}$ in the air boundary layer. 

We define the perturbed part of dimensionless heat flux from the ice-water interface to the water and from the ice to the ice-water interface, as  
$q_{l*}\equiv \Imag [-\partial T'_{l*}/\partial y_{*}|_{y_{*}=\zeta_{*}}]$ and
$q_{s*}\equiv \Imag[-K^{s}_{l}\partial T'_{s*}/\partial y_{*}|_{y_{*}=\zeta_{*}}]$, respectively,
where $T'_{l*}$ and $T'_{s*}$ represent the perturbed terms in Eqs. (\ref{eq:Tl}) and (\ref{eq:Ts}).
Hence, the total heat flux from the ice-water interface to the water and ice is expressed as follows: \cite{Ueno09}
\begin{eqnarray}
q_{ls*} &\equiv& q_{l*}-q_{s*}
=\delta_{b}\Imag\left[\left\{-\frac{dH_{l}}{dy_{*}}\Big|_{y_{*}=0}+K^{s}_{l}\mu_{l}(H_{l}|_{y_{*}=0}-1)\right\}
                 {\rm exp}(\sigma_{*}t_{*}+i\mu_{l}x_{*})\right] \nonumber \\
&=&\left[\left\{-\frac{dH_{l}^{(r)}}{dy_{*}}\Big|_{y_{*}=0}+K^{s}_{l}\mu_{l}(H_{l}^{(r)}|_{y_{*}=0}-1)\right\}^{2} 
+\left\{-\frac{dH_{l}^{(i)}}{dy_{*}}\Big|_{y_{*}=0}+K^{s}_{l}\mu_{l}H_{l}^{(i)}|_{y_{*}=0}\right\}^{2}\right]^{1/2} \nonumber \\
&&\times \delta_{b}(t_{*})\sin[\mu_{l}(x_{*}-v_{p*}t_{*})-\Theta_{q_{ls*}}],
\label{eq:ql-qs}
\end{eqnarray}
where $\Theta_{q_{ls*}}$ is a phase difference between the total heat flux $q_{ls*}$ at $y_{*}=\zeta_{*}$ and the ice-water interface. 
We also define the perturbed part of dimensionless heat flux from the water-air surface to the air as $q_{a*}\equiv \Imag[-\partial T'_{a*}/\partial \eta|_{\eta=\xi'/\delta_{0}}]$,
where $T'_{a*}=T'_{a}/(T_{la}-T_{\infty})$ represents the perturbed term in Eq. (\ref{eq:Ta}). Hence,
\begin{eqnarray}
q_{a*}&=&-\frac{\zeta_{k}}{\delta_{0}}
\Imag\left[\frac{dH_{a}}{d\eta}\Big|_{\eta=0}f_{l}|_{y_{*}=1}{\rm exp}(\sigma_{*}t_{*}+i\mu_{l}x_{*})\right] \nonumber \\
&=&
\left[\left(G'^{(r)}_{a}f_{l}^{(r)}|_{y_{*}=1}-G'^{(i)}_{a}f_{l}^{(i)}|_{y_{*}=1}\right)^{2} 
+\left(G'^{(r)}_{a}f_{l}^{(i)}|_{y_{*}=1}+G'^{(i)}_{a}f_{l}^{(r)}|_{y_{*}=1}\right)^{2}\right]^{1/2} \nonumber \\ 
&& \times \delta_{b}(t_{*})\sin[\mu_{l}(x_{*}-v_{p*}t_{*})-\Theta_{q_{a*}}], 
\label{eq:qa}
\end{eqnarray}
where $\Theta_{q_{a*}}$ is a phase difference between the heat flux $q_{a*}$ at $\eta=\xi'/\delta_{0}$ and the ice-water interface. 

\begin{figure}[ht]
\begin{center}
\includegraphics[width=7cm,height=7cm,keepaspectratio,clip]{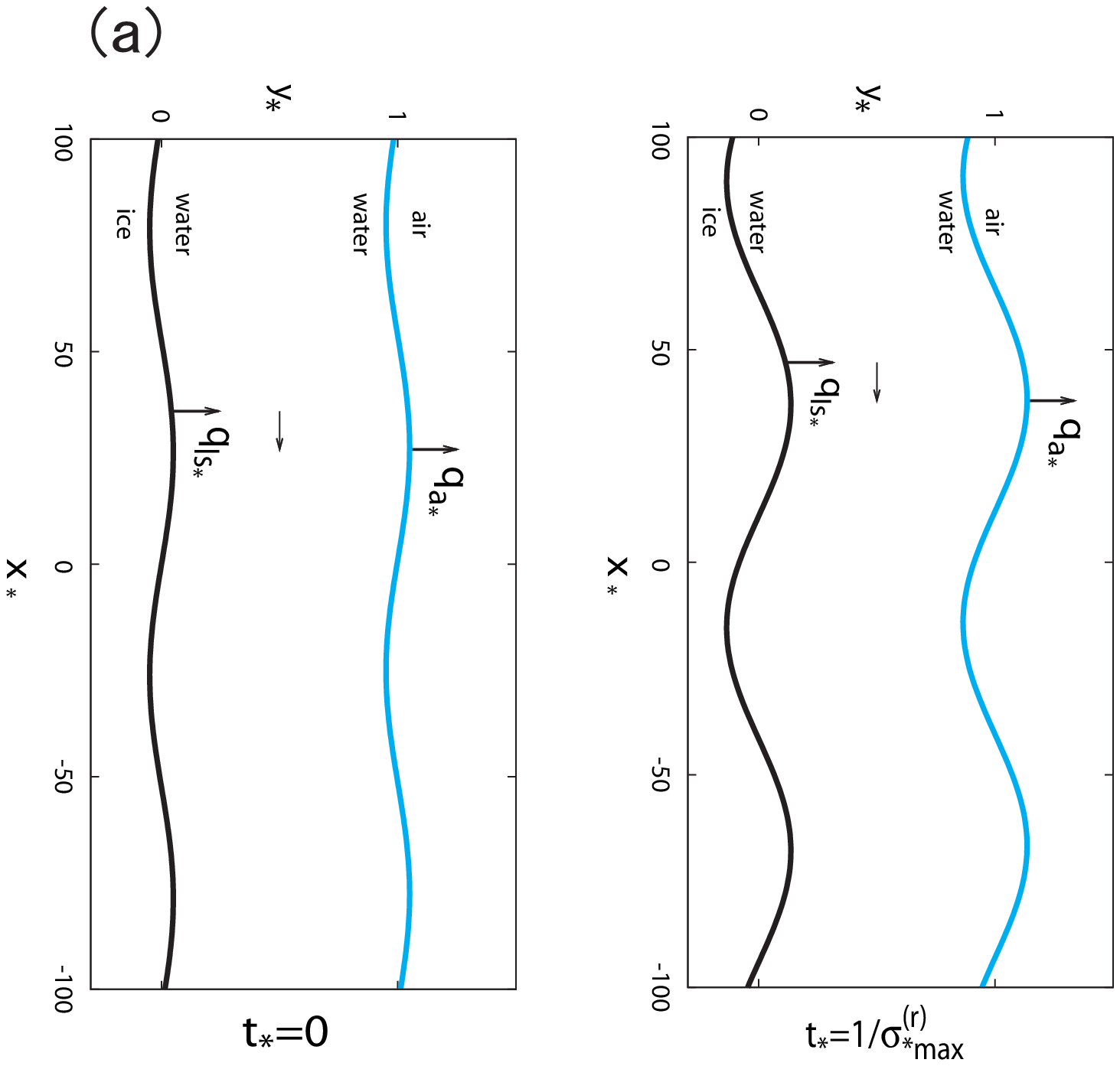}\hspace{5mm}
\includegraphics[width=7cm,height=7cm,keepaspectratio,clip]{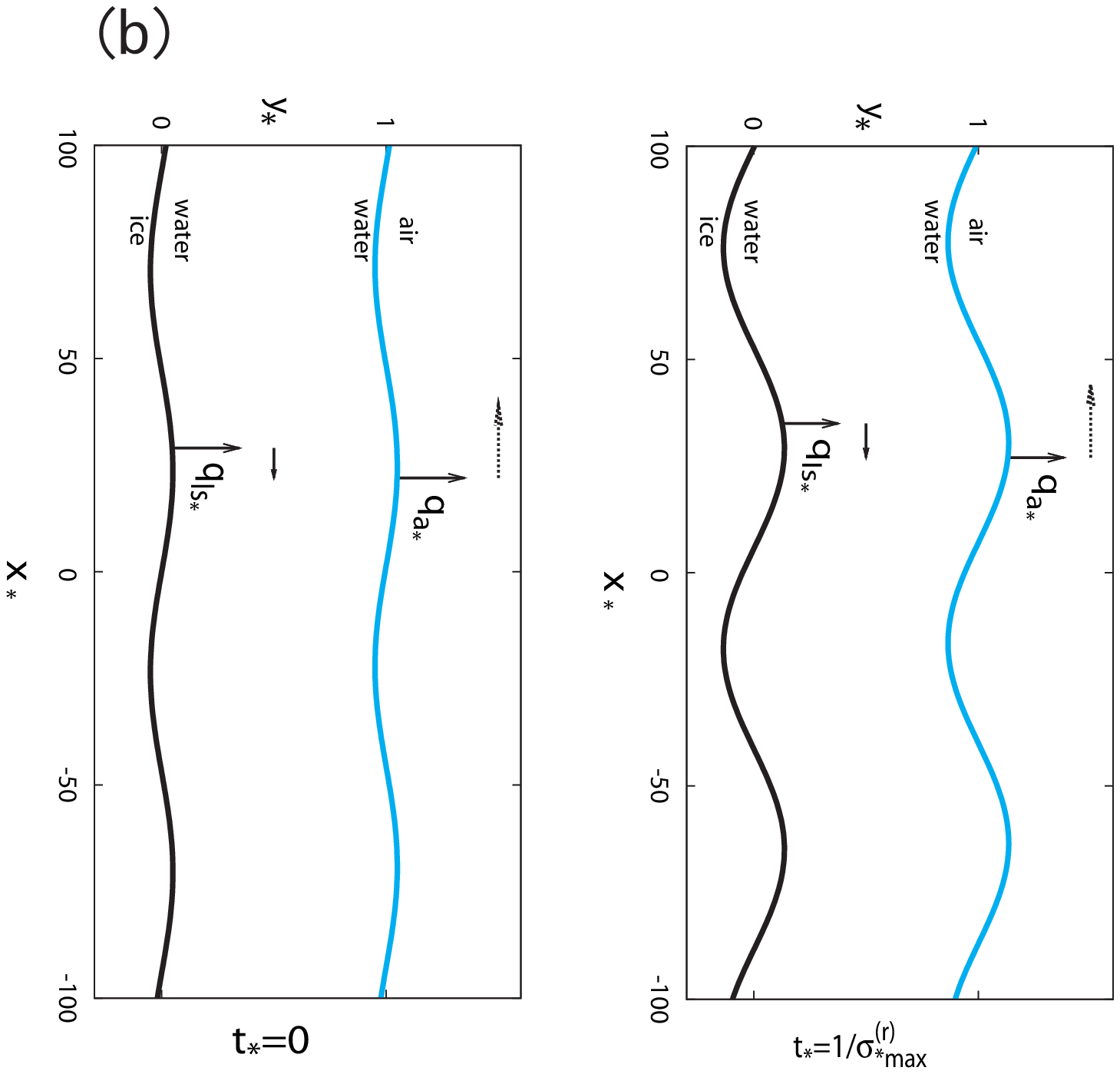}\\[2mm]
\includegraphics[width=8cm,height=8cm,keepaspectratio,clip]{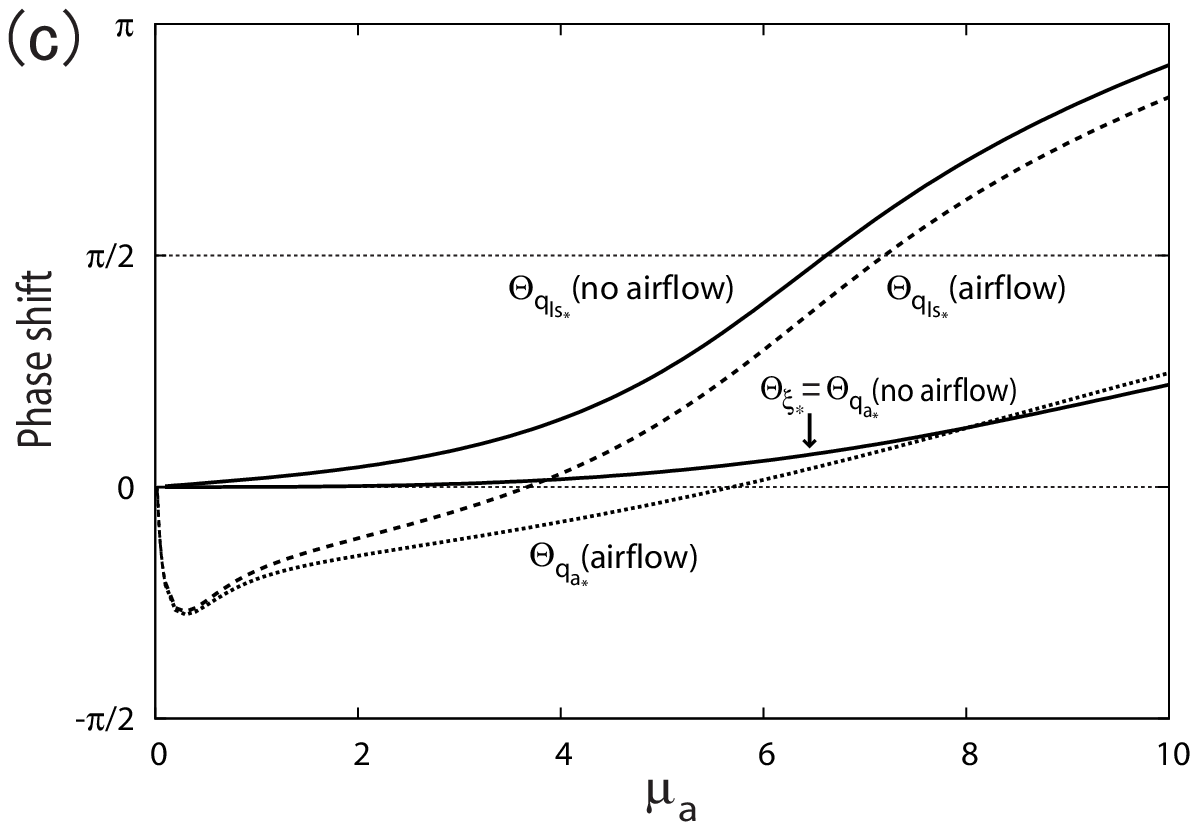}
\end{center}
\caption{(a) and (b) are illustrations of the time evolution of an initial disturbance of the ice-water interface from $t_{*}=0$ to $t_{*}=1/\sigma^{(r)}_{*\rm max}$. The solid arrows in the water film and the dashed arrows in the air boundary layer show the direction of the supercooled water flow and airflow, respectively. The arrows attached $q_{la*}$ and $q_{a*}$ are the maximum point of heat flux at the ice-water interface and water-air surface, respectively.
(a) represents the disturbance of $\mu_{a}=4.3$ in the absence of airflow. 
(b) represents the disturbance of $\mu_{a}=4.8$ in the presence of airflow. 
(c) represents the phase shift of the water-air surface, $\Theta_{\xi_{k*}}$, 
total heat flux at the ice-water interface, $\Theta_{q_{ls*}}$, 
and heat flux at the water-air surface, $\Theta_{q_{a*}}$, 
relative to the ice-water interface with respect to $\mu_{a}$.}
\label{fig:heatflux-sla-phase}
\end{figure}

Figures \ref{fig:heatflux-sla-phase} (a) and (b) illustrate the time evolution of the ice-water interface with an initial amplitude of $\delta_{b}=0.05$, for the wave number $\mu_{a}=4.3$ in the absence of airflow and for $\mu_{a}=4.8$ in the presence of airflow, respectively. The respective wave number represents the fastest growing mode, at which $\sigma^{(r)}_{*}$ acquires a maximum value, as shown by the dashed and solid lines in Fig. \ref{fig:sim-mua-amp-vp} (a). 
The arrows on the ice-water interface and the water-air surface show the position of the maximum of $q_{ls*}$
and that of $q_{a*}$.
Using Eq. (\ref{eq:dHlr-dHli-airflow}), Eq. (\ref{eq:qa}) can be written as
$q_{a*}=[(-dH^{(r)}_{l}/dy_{*}|_{y_{*}=1})^{2}+(-dH^{(i)}_{l}/dy_{*}|_{y_{*}=1})^2]^{1/2}
\delta_{b}(t_{*})\sin[\mu_{l}(x_{*}-v_{p*}t_{*})-\Theta_{q_{a*}}]$. 
Therefore, non-zero values of $-dH^{(i)}_{l}/dy_{*}|_{y_{*}=1}$ in Eq. (\ref{eq:dHlr-dHli-airflow}) contribute to the imaginary part of $q_{a*}$, and cause the phase shift of $q_{a*}$ relative to the ice-water interface. 

In the absence of airflow, as shown in  Fig. \ref{fig:heatflux-sla-phase} (a), $q_{a*}$ is largest at each protruded part of the water-air surface because the isotherm in the air is symmetrical around the protruded part. As shown in  Fig. \ref{fig:heatflux-sla-phase} (c), the water-air surface shifts to the positive $x_{*}$ direction by $\Theta_{\xi_{*}}$ relative to the ice-water interface. In the absence of airflow, since $G'^{(r)}_{a}=\mu_{l}$ and $G'^{(i)}_{a}=0$, Eq. (\ref{eq:qa}) yields 
$q_{a*}=\mu_{l}[(f_{l}^{(r)}|_{y_{*}=1})^{2}+(f_{l}^{(i)}|_{y_{*}=1})^2]^{1/2}
\delta_{b}(t_{*})\sin[\mu_{l}(x_{*}-v_{p*}t_{*})-\Theta_{q_{a*}}]$. 
Comparing this to Eq. (\ref{eq:xi}), it is found that the position of the maximum of $q_{a*}$ also shifts to the positive $x_{*}$ direction by $\Theta_{q_{a*}}=\Theta_{\xi_{*}}$. 
However, the position of the maximum of $q_{ls*}$ shifts by $\Theta_{q_{ls*}}$ to the upper side of the protruded part of the ice-water interface. 

On the other hand, in the presence of an upward airflow shown by the dashed arrow in Fig. \ref{fig:heatflux-sla-phase} (b), the isotherms in the air boundary layer are no longer symmetrical around each protruded part.
The isotherms become closer on the lower side of the protruded part of the water-air surface due to the upward airflow. Hence, $q_{a*}$ is largest on the lower side of the protruded part, as shown in  Fig. \ref{fig:heatflux-sla-phase} (b). 
By comparing Fig. \ref{fig:heatflux-sla-phase} (a) to (b), 
first, it is found that the position of the maximum of $q_{a*}$ in the absence of airflow is always on the protruded part of the water-air surface, but that position changes by the presence of airflow and depends on the wave number $\mu_{a}$.
As shown in Fig. \ref{fig:heatflux-sla-phase} (c), the sign of $\Theta_{q_{a*}}$ in the presence of airflow changes from negative to positive value at $\mu_{a}=5.7$, which corresponds to the change of sign of $-dH^{(i)}_{l}/dy_{*}|_{y_{*}=1}$ in Fig. \ref{fig:sim-mua-amp-vp} (c).  
Second, there is a critical difference between the phase shift $\Theta_{q_{ls*}}$ in the absence of airflow and that in its presence. In the absence of airflow, the position of the maximum of $q_{ls*}$ shifts to the upper side of the protruded part of the ice-water interface with an increase in $\mu_{a}$ (see $\Theta_{q_{ls*}}$ (no airflow) in Fig. \ref{fig:heatflux-sla-phase} (c)). In this case, the sign of $v_{p*}$ is positive as shown by the dashed line in Fig. \ref{fig:sim-mua-amp-vp} (b). 
Figure \ref{fig:heatflux-sla-phase} (a) shows that the ripple at $\mu_{a}=4.3$ moves upwards at $v_{p*}=0.59$. The displacement in the dimensional form is about 11 $h_{0}$ after the dimensionless time $1/\sigma^{(r)}_{*\rm max}=1/0.054$.   
On the other hand, in the presence of upward airflow, the position of the maximum of $q_{ls*}$ is on the lower side of the protruded part of the ice-water interface for $0<\mu_{a}<3.7$, whereas that is on the upper side for $\mu_{a}>3.7$ (see $\Theta_{q_{ls*}}$ (airflow) in Fig. \ref{fig:heatflux-sla-phase} (c)). We showed that the sign of $v_{p*}$ changes from negative to positive at $\mu_{a}=3.7$ by the solid line in Fig. \ref{fig:sim-mua-amp-vp} (b). 
Therefore, the sign of $v_{p*}$ is related to the sign of $\Theta_{q_{ls*}}$.
The ripples move down in the mode $0<\mu_{a}<3.7$, whereas they move up in the mode $\mu_{a}>3.7$. However, the ripple with the most unstable mode of $\mu_{a}=4.8$ is expected to be observed. Figure \ref{fig:heatflux-sla-phase} (b) shows that the ripple at $\mu_{a}=4.8$ moves upwards at $v_{p*}=0.48$. The displacement in the dimensional form is about 6 $h_{0}$ after the dimensionless time $1/\sigma^{(r)}_{*\rm max}=1/0.085$.  

Substituting Eqs. (\ref{eq:Tl}) and (\ref{eq:Ts}) into Eq. (\ref{eq:Tsl}), the dimensionless form of $\Delta T_{sl}$ can be written as:
\begin{equation}
\Delta T_{sl*}
=[(H_{l}^{(r)}|_{y_{*}=0}-1)^{2}+(H_{l}^{(i)}|_{y_{*}=0})^{2}]^{1/2}
\delta_{b}(t_{*})\sin[\mu_{l}(x_{*}-v_{p*}t_{*})-\Theta_{T_{\zeta_{*}}}],
\label{eq:DeltaTsl}  
\end{equation}
where $\Theta_{T_{\zeta_{*}}}$ is a phase difference between the temperature at $y_{*}=\zeta_{*}$ and the ice-water interface. $H_{l}^{(r)}$ and $H_{l}^{(i)}$ in (\ref{eq:DeltaTsl}) are determined by solving (\ref{eq:geq-Hl}) with the boundary conditions (\ref{eq:Tla-xi}) and (\ref{eq:heatflux-xi-h0}). Since the water film flow is not affected by the natural convection airflow, the forms of $\bar{U}_{l*}|_{y_{*}=1}$ and $f_{l}|_{y_{*}=1}$ in (\ref{eq:Tla-xi}) and (\ref{eq:heatflux-xi-h0}) are the same as those in the absence of airflow. However, $dH_{a}/d\eta|_{\eta=0}$ in (\ref{eq:heatflux-xi-h0}) in the presence of airflow is different from that in the absence of airflow.
As a result of the change in the temperature gradient at the water-air surface, the solution $H_{l}$ changes and causes different distribution of $\Delta T_{sl*}$ at the ice-water interface. The position of the maximum of $\Delta T_{sl*}$ changes depending on that of the rate of latent heat loss at the water-air surface. \cite{Ueno09} That is why $\Delta T_{sl}$ in Eq. (\ref{eq:Tsl}) was considered as the spatial temperature non-uniformity  caused by the external disturbance at the water-air surface. Heat flux $q_{s*}$ in the ice in the vicinity of the ice-water interface is caused due to the deviation $\Delta T_{sl*}$, which contributes to the second term in Eq. (\ref{eq:dispersion}).

\section{Summary and Discussion}

A morphological instability theory has been elaborated for ice growth under a water film flow with a free surface and a natural convection airflow, within a linear stability analysis. This theory proposes a synthetic treatment of the heat flow in ice, water and air through a disturbed ice-water interface and water-air surface, thin water film flow and airflow, taking into account the influence of the shape of the water-air surface on the growth condition of the ice-water interface disturbance.
Even though the natural convection airflow was introduced, the shear stress-free condition at the unperturbed water-air surface still held. Moreover, the influence of the perturbed part of shear and normal stresses due to natural convection airflow on the water film flow was negligible. Consequently, the perturbed distribution of water film flow could be obtained without considering the influence of the airflow. However, since the rate of latent heat loss from the water-air surface to the surrounding air is affected by the airflow, the perturbed temperature distribution in the water layer is different from that in the absence of airflow.
In the absence of airflow, the position of the maximum of heat flux $q_{a*}$ at the water-air surface is at the protruded part of the water-air surface. In the presence of airflow, that of $q_{a*}$ is not necessarily at the protruded part.  
Depending on the position of the maximum of $q_{a*}$, that of $q_{ls*}$ at the ice-water interface changes. We find that the position of the maximum growth rate of the ice-water interface disturbance is shifted upward relative to the position of the maximum of $q_{a*}$. 
We also find that although the airflow causes the amplification rate of the ice-water interface disturbance to increase by the enhancement of the rate of latent heat loss from the water-air surface to the surrounding air, the wavelength of ice ripples is not significantly affected by the natural convection airflow.
On the other hand, the mean ice growth rate $\bar{V}$ and the ripple translation velocity $v_{p}$ depend on the parameters characterizing the air boundary layer. 

We mention the importance of the influence of the temperature distribution in water film flow on the growth condition of the ice-water interface disturbance even though the water layer is very thin. If we can neglect the temperature distribution within the water layer, and focus on only the temperature distribution in the air, Eq. (\ref{eq:heatflux-zeta}) is replaced by $L(\bar{V}+\partial \zeta/\partial t)=-K_{a}\partial T_{a}/\partial y|_{y=\xi}$. Linearizing this equation at $y=h_{0}$ yields, to the zeroth order in $\xi_{k}$, $\bar{V}=-K_{a}T_{\infty}/(L\delta_{0}/\bar{G}_{a*})$,
which is identical to Eq. (\ref{eq:V-airflow}).
The first order in $\xi_{k}$ gives
$\sigma=(\bar{V}/h_{0})(h_{0}/\delta_{0})(-dH_{a}/d\eta|_{\eta=0})f_{l}|_{y_{*}=1}$, 
whose real part is approximately expressed as
$\sigma_{*}^{(r)}
=G'^{(r)}_{a}f_{l}^{(r)}|_{y_{*}=1}-G'^{(i)}_{a}f_{l}^{(i)}|_{y_{*}=1}
=(36G'^{(r)}_{a}+6\alpha G'^{(i)}_{a})/(36+\alpha^{2}) \approx G'^{(r)}_{a}$.
Comparing this to Eq. (\ref{eq:amp-approx-airflow}), it is found from Fig. \ref{fig:tempprofiles-Ga} (b) that the ice-water interface is always unstable because the stabilizing term is absent. It should be noted that the stabilizing term in Eq. (\ref{eq:amp-approx-airflow}) was obtained from the solution of the perturbed temperature distribution in the water film flow. Although the heat transfer through the air boundary layer is the deciding factor in the growth rate $\bar{V}$, in order to obtain the growth condition of the ice-water interface disturbance, it is important to determine the perturbed temperature distribution in the water layer as well as that in the air boundary layer. 

Although the wavelengths theoretically obtained in Table \ref{tab:tableI} are in agreement with experimental results, \cite{Maeno94, Matsuda97, Ueno09} several questions arise for the values of $\bar{V}$ and $T_{la}$. The measured growth rates of icicle radius in the experiment was about $1.4 \sim 5.3$ mm/h for the air temperatures range of $-4.9 \sim -28.8$ $^\circ$C in the case of the zero wind speed. \cite{Maeno94}  
Also, the mean radial growth rate of ice grown on a 6-mm diameter round stick was 1.7 mm/h (see Figure 9 (a) \cite{Ueno09}). This experiment was conducted in a cold room, where large temperature fluctuations of $\pm3$ $^\circ$C  around $-9$ $^\circ$C were observed. Substituting the measured value into the energy conservation equation $L\bar{V}=-K_{l}T_{la}/h_{0}$ at the ice-water interface, the degree of supercooling of the water layer is $T_{la}=-0.03$ $^\circ$C. Certainly, the values of $T_{la}$ and $\bar{V}$ calculated from Eqs. (\ref{eq:Tla-airflow}) and (\ref{eq:V-airflow}) are less than the measured experimental values by one order of magnitude.
If the value of the boundary layer thickness $\delta$ is less than that estimated from the natural convection boundary layer, Eqs. (\ref{eq:Tla-airflow}) and (\ref{eq:V-airflow}) suggest that the values of $T_{la}$ and $\bar{V}$ should increase. It is known that it is somewhat difficult to grow icicles with significant ripples in the steady calm conditions of icicle formation. \cite{Maeno94} Therefore, instead of assuming a calm environment for ice growth, different heat transfer mechanisms needs to be considered. Also, it is necessary to predict or to measure the mean ice growth rate $\bar{V}$ accurately in order to estimate the displacement of ripples.
We have to be careful in the measurement of the displacement of ripples because $v_{p}$ depends on environmental conditions, as mentioned above.

Finally, limitations of the proposed theory must be mentioned. 
First, it was assumed that ice was grown in a flat gutter on an inclined plane, considering a perturbation around the flat ice surface. 
However, as shown in Table \ref{tab:tableI}, for a given air temperature $T_{\infty}$, since the ice growth rate $\bar{V}$ depends on $x$, the actual grown ice thickness on the gutter varies locally. If heat conduction through the ice to the substrate is negligible, the ice thickness $b_{0}$ in the unperturbed state is proportional to the time. \cite{Ueno09} The angle that tangent vector to the ice-water interface at $(x,b_{0})$ makes with respect to the positive $x$ direction is given by $\phi(x,t)=\cos^{-1}[\{1+(db_{0}/dx)^{2}\}^{-1/2}]$.
Making use of Eq. (\ref{eq:V-airflow}), $x=h_{0}x_{*}$ and $t=(h_{0}/\bar{V})t_{*}$, 
$\phi(x,t)$ gradually changes in time from the initial flat ice surface by 
$\cos^{-1}[\{1+\{(d\bar{V}/dx)t\}^{2}\}^{-1/2}]
=\cos^{-1}[\{1+\{t_{*}/(4x_{*})\}^{2}\}^{-1/2}]$.
However, the change is negligible except for small $x_{*}$ because $t_{*}/x_{*} \ll 1$ even after 10 hours in the range of $0.1 \leq x \leq 1$ m. The actual geometry of the icicle is that of an elongated carrot shape. \cite{Short06} In this case too, since icicle's surfaces are nearly vertical, we can neglect the change in the slope $db_{0}/dx$ in $\phi(x,t)$ except for the tip region.
Hence, the use of air boundary layer under the assumption of a flat ice surface is valid, \cite{Short06} and the local variation in the thickness $h_{0}$ and the surface velocity $u_{l0}$ of the water film in the unperturbed state is negligible as ice grows.
However, for ice growth on aircraft wings and aerial cables, the local change in $\phi$ in time is remarkable compared to the icicle growth, so that we have to consider a morphological instability around curved ice surfaces in the unperturbed state, and $h_{0}$ and $u_{l0}$ are no longer constant over the curved ice surface. This is relevant to the problems on solidification on surfaces of arbitrary curvature. \cite{Myers02}  

Second, in our linear stability analysis, a small perturbation of the ice-water interface was assumed: 
$y_{*}=\zeta_{*}=\delta_{b}(t_{*})\sin[\mu_{l}(x_{*}-v_{p*}t_{*})]$. 
However, since the amplitude $\delta_{b}(t_{*})=\delta_{b}{\rm exp}(\sigma^{(r)}_{*}t_{*})$ in $\zeta_{*}$ and in the corresponding fields increases exponentially with time when $\sigma^{(r)}_{*}>0$, the non-linear terms for the perturbation in the governing equations and boundary conditions are no longer  small. 
Even though the linear approximation only describes the initial evolution of the perturbation, there was good agreement between the wavelengths predicted from our linear stability analysis and experimentally observed wavelengths of finite amplitude ripples. However, it is needless to say that the linear theory is unable to clarify further features related to ripple development, and the question arises of the value of the saturation amplitude, and of how the perturbation amplitude evolves towards this value. \cite{Caroli92} This leads us to extend the linear perturbation calculation to higher orders in the perturbation amplitude. \cite{Wollkind70} Such an amplitude expansion generalizes the time evolution equation of the amplitude of the ice-water interface from  $d\delta_{b}(t_{*})/dt_{*}=\sigma^{(r)}_{*}\delta_{b}(t_{*})$ to a nonlinear amplitude evolution equation. In order to implement it, algebraically complicated calculations are needed.

Third, for the relatively weak flow considered here, the free shear stress condition at the water-air surface was still satisfied, and water film flow was driven by gravity only. However, in the presence of a strong airflow around aircraft wings and aerial cables, the water film flow is driven by gravity and aerodynamic forces. Due to strong air shear stress exerted on the water-sir surface, the distribution of water film flow must be modified from the half-parabolic form $\bar{U}_{l*}=y_{*}^{2}-2y_{*}$ to $\bar{U}_{l*}=y_{*}^{2}+(R_{\tau_{al}}-2)y_{*}$, as discussed in \ref{sec:linearization}.
It is known that the aerodynamic forces, as modified by the accreted ice, are significant in determining the wind drag and lift on iced structures. However, the traditional approach in wet icing modelings has been based on the mass and energy conservations only and have ignored the dynamics of the surface flow of unfrozen water. \cite{Farzaneh08} When airflow and water film flow are coupled, the distribution of shear and normal stresses at the water-air surface may influence the temperature distribution in the water layer. The action of an aerodynamic force on the water-air surface, and the resulting morphological instability of the ice-water interface have to be considered. These issues are beyond the scope of the analysis developed here.
Removing these restrictions will be the subject of future research.

\begin{acknowledgements}
This study was carried out within the framework of the NSERC/Hydro-Qu$\acute{\rm e}$bec/UQAC Industrial Chair on Atmospheric Icing of Power Network Equipment (CIGELE) and the Canada Research Chair on Engineering of Power Network Atmospheric Icing (INGIVRE) at the Universit$\acute{\rm e}$ du Qu$\acute{\rm e}$bec $\grave{\rm a}$ Chicoutimi. 
The authors would like to thank all CIGELE partners (Hydro-Qu$\acute{\rm e}$bec, Hydro One, R$\acute{\rm e}$seau Transport d'$\acute{\rm E}$lectricit$\acute{\rm e}$ (RTE) and $\acute{\rm E}$lectricit$\acute{\rm e}$ de France (EDF), Alcan Cable, K-Line Insulators, Tyco Electronics, Dual-ADE, and FUQAC) whose financial support made this research possible. 
\end{acknowledgements}


\begin{thebibliography}{99}

\bibitem{Maeno94}
N. Maeno, L. Makkonen, K. Nishimura, K. Kosugi and T. Takahashi,
``Growth rates of icicles,"  
J. Glaciol \textbf {40}, 319 (1994).

\bibitem{Matsuda97}
S. Matsuda, 
``Experimental study on the wavy pattern of icicle's surface,"
M.S. thesis, Hokkaido University (1997).

\bibitem{Ogawa02}
N. Ogawa and Y. Furukawa,
``Surface instability of icicles,"
Phys. Rev. E \textbf {66}, 041202 (2002).

\bibitem{Ueno03}
K. Ueno, 
``Pattern formation in crystal growth under parabolic shear flow,"
Phys. Rev. E \textbf{68}, 021603 (2003).

\bibitem{Ueno04}
K. Ueno, 
``Pattern formation in crystal growth under parabolic shear flow II,"
Phys. Rev. E \textbf{69}, 051604 (2004).

\bibitem{Ueno07}
K. Ueno,
``Characteristics of the wavelength of ripples on icicles," 
Phys. Fluids \textbf{19}, 093602 (2007).

\bibitem{Ueno09}
K. Ueno, M. Farzaneh, S. Yamaguchi, and H. Tsuji,
``Numerical and experimental verification of a theoretical model of ripple formation 
in ice growth under supercooled water film flow,"
Fluid Dyn. Res. \textbf{42}, 025508 (2010). 

\bibitem{Short06}
M. B. Short, J. C. Baygents, and R. E. Goldstein,
``A free-boundary theory for the shape of the ideal dripping icicle,"
Phys. Fluids \textbf{18}, 083101 (2006).

\bibitem{Benjamin57}
T. B. Benjamin,
``Wave formation in laminar flow down an inclined plane,"
J. Fluid Mech \textbf{2}, 554 (1957). 

\bibitem{Landau59}
L. Landau and E. Lifschitz, 
\textit{Fluid Mechanics}
(Pergamon Press, London, 1959).

\bibitem{Craik66}
A. D. D. Craik,
``Wind-generated waves in thin liquid films,"
J. Fluid Mech \textbf{26}, 369 (1966). 

\bibitem{Yih67}
C.-S. Yih,
``Instability due to viscosity stratification,"
J. Fluid Mech \textbf{27}, 337 (1967). 

\bibitem{Butler02}
S. Butler and P. Harrowell, 
``Factors determining crystal-liquid coexistence under shear,"
Nature \textbf{145}, 1008 (2002).

\bibitem{Caroli92}
B. Caroli, C. Caroli and B. Roulet 
``Instabilities of planar solidification fronts,"
in \textit{Solids Far From Equilibrium},
edited by Godr$\grave{\rm e}$che C (Cambridge University Press, Cambridge, 1992)

\bibitem{Gebhart73}
B. Gebhart,
``Instability, transition, and turbulence in buoyancy induced flows,"
Annual Review of Fluid Mechanics \textbf{15}, 213 (1973).

\bibitem{Schlichting99}
H. Schlichting and K. Gersten, 
\textit{Boundary Layer Theory}
(Springer, Berlin, 1999).

\bibitem{Myers02}
T. G. Myers, J. P. F. Charpin and S. J. Chapman,
``The flow and solidification of a thin fluid film on an arbitrary three-dimensional surface",
Phys Fluids \textbf{14}, 2788 (2002). 

\bibitem{Wollkind70}
D. J. Wollkind and L. A. Segel,
``A nonlinear stability analysis of the freezing of a dilute binary alloy,"
Phil. Trans. R. Soc. Lond. A \textbf{268}, 351 (1970).

\bibitem{Farzaneh08}
L. Makkonen and E. P. Lozowski,
``Numerical modelling of icing on power network equipment,"
in \textit{Atmospheric Icing of Power Networks},
edited by M. Farzaneh
(Springer, Berlin, 2008).

\end{thebibliography}
\end{document}